\documentclass[preprint,12pt]{elsarticle}

\usepackage{url}
\usepackage{setspace}
\usepackage{ctable}
\usepackage[latin1]{inputenc}
\usepackage{pslatex} %makes times new roman
\usepackage{natbib}
\usepackage{amsmath}
\usepackage[normalem]{ulem}
\usepackage{amssymb}
\usepackage[flushleft]{threeparttable}
\usepackage{ctable}
\usepackage{bm}
\usepackage{graphicx}
\usepackage{cancel}
\usepackage{array}
\usepackage{color}
\usepackage{colortbl}
\usepackage{pdflscape}
\usepackage{afterpage} 
\usepackage{float}
\usepackage{multirow} % consente multi righe nelle tabelle
\usepackage{subfigure} % crea sottotabelle e sottofigure
\usepackage{rotating}
\usepackage{bbm}
\usepackage{amssymb}
\usepackage{amsthm}

\usepackage{chngcntr}
\usepackage{apptools}
\usepackage[margin=1.5in]{geometry}
\newtheorem{thm}{Theorem}
\newtheorem{cor}{Corollary}

\usepackage{lineno}

\journal{Journal}

\begin{document}

\begin{frontmatter}



\title{Modeling Portfolios with Leptokurtic and Dependent Risk Factors}
\author[ua]{Piero Quatto}
\ead{piero.quatto@unimib.it}
\author[ub]{Gianmarco Vacca}
\ead{gianmarco.vacca@unicatt.it}
\author[ub]{Maria Grazia Zoia\corref{corr}}
\ead{maria.zoia@unicatt.it, phone: +390272342948}
\address[ub]{Universit\`{a} Cattolica del Sacro Cuore, Largo Gemelli 1, 20123 Milano, Italy}
\address[ua]{Universit\`{a} degli Studi di Milano Bicocca, Piazza dell'Ateneo Nuovo 1, 20126 Milano, Italy}
\cortext[corr]{Corresponding Author}

\begin{abstract}
Recently, an approach to modeling portfolio-distribution with risk factors distributed as Gram-Charlier (GC) expansions of the Gaussian law, has been conceived. GC expansions prove effective when dealing with moderately leptokurtic data. In order to cover the case of possibly severe leptokurtosis, the so-called GC-like expansions have been devised by reshaping parent leptokurtic distributions by means of orthogonal polynomials specific to them. In this paper, we focus on the hyperbolic-secant (HS) law as parent distribution whose GC-like expansions fit with kurtosis levels up to 19.4. A portfolio distribution has been obtained with risk factors modeled as GC-like expansions of the HS law which duly account for excess kurtosis. 
%Given that dependence, in the form of between-square correlation, occurs frequently, a copula function has been worked out on an orthogonal-polynomial argument, with the portfolio distribution recast accordingly. 
Empirical evidence of the workings of the approach dealt with in the paper is included. 

\end{abstract}

\begin{keyword}

 Gram-Charlier-like expansions \sep Orthogonal polynomials\sep Kurtosis \sep Value at Risk \sep Expected Shortfall.

\end{keyword}
%\Jel Classifications: C1 \sep G1 
\end{frontmatter}

\section{Introduction}\label{sec:intro}
Recently, Zoia et al. \cite{Zoia2018} proposed a new approach to model the distribution of a portfolio which hinges on the representation of its returns or insurance losses as Gram-Charlier (GC) expansions. The resulting portfolio distribution was proved to be tail sensitive and, as such, suitable for computing risk measures like the Value at Risk (VaR) and the Expected Shortfall (ES). However, the Gram-Charlier expansion of the Gaussian law based on Hermite orthogonal polynomials is fit only for series with moderate excess kurtosis  (lower than 5). Since all too often financial series exhibit possibly severe kurtosis, an orthogonal-polynomial technique has been also worked out to cover leptokurtic distributions \citep{Bagnato2015a}. This has led to the class of Gram-Charlier-like (GCl) expansions. 
%Gram-Charlier-like (GCl) expansions can be used in their stead \sout{the modeling of a portfolio's returns should be done by using Gram-Charlier-like (GCl) expansions, that is}, being polynomial expansions of already  leptokurtic distributions \sout{which are able to account for high kurtosis}. 
In this paper, we examine the GCl expansions of the hyperbolic secant distribution (HS). This type of distribution can be traced back to \cite{Fisher1921}, \cite{Dodd1925}, \cite{Roa1924} and \cite{Perks1932}. Although it is the generator distribution of the sixth natural exponential family (NEF) with quadratic variance (see \cite{Morris1982} and \cite{Morris2009}), HS is somewhat less known than other distributions of the same family (see \cite{Ding2014a}). HS is a symmetric and bell-shaped distribution, like the Gaussian law, with cumulative and quantile functions that have simple closed-form expressions which makes it appealing for practical purposes. Unlike stable distributions, the HS law has finite moments of every order that can be conveniently expressed in terms of the Euler numbers. More importantly, being itself leptokurtic (its kurtosis is equal to 5), it is an appealing parent law for GCl expansions able to fit 
%\sout{it is an appealing starting distribution to get, via its polynomial expansion, distributions which exhibit tails heavier than those attainable via GC expansions. In \cite{Bagnato2015a} it has proved that the polynomially adjusted hyperbolic secant can be used to fit 
empirical distributions with severe kurtosis up to 19.4 \citep[see][]{Bagnato2015a}. Thus, it proves useful to modeling series exhibiting fat tails \cite{Vaughan2002}, such as portfolio returns. \\
In this paper, we obtain the density of the sum of GCl expansions of independent HS laws with equal or different kurtoses (SGCHS) by using Fourier transform techniques. 
%by working out the relationships between the Fourier Transform (FT) and the corresponding anti-Fourier Transforms of the HS law which, as it is well known, tally with the characteristic and density functions of a random variable, the HS in this particular case. 
%%and its relation to characteristic functions of the HS and GCHS \sout{which tally with the characteristic functions of} random variables \sout{whose densities are the associated anti-Fourier transforms.}.  
Furthermore, the  between square dependance, which usually plagues financial data, has been duly accounted for thanks to a copula of new conception, whose rationale rests on an orthogonal polynomial argument. This copula has been designed for the type of dependence, known as between-squares correlation, which is likely to be found when dealing with volatility with thick tails.
% \sout{as the extensive use of ARCH and GARCH models proves \cite{Bollerslev1987} \cite{Harvey1999} \cite{Nelsen2007} \cite{Szego2004}}. The idea behind this copula is to link margins by embodying information about between-square correlation via orthogonal polynomials. \\
The performance of SGCHS distributions, tailored to account for between-square dependence, in modeling portfolios has been assessed with an empirical application on a portfolio composed of some financial series. 
%\sout{Density functions of the sums of GC-like expansions of HS laws encoding between-square dependence have been used to model portfolios composed of the said indexes and risks measures such as the value at risk} 
Both in sample and out-of-sample VaR and ES values have been obtained in order to test the performance of SGCHS distributions in estimating and predicting these risk measures.\\
Both a Maximum Likelihood procedure, known as Inference for the Margins (IFM), and the method of moments have been employed to fit SGCHS to the data. Also specific tests have been carried out to evaluate their in and out-of-sample performance in evaluating risk measures. This empirical analysis provides evidence that modeling asset returns with GC-like expansions of HS laws and accounting for marginal dependence via the proposed copula leads to portfolio densities which better capture risk and expected losses, especially for low confidence levels.\\  
The paper is organized as follows. Section \ref{sec:theory} presents the theoretical results on the SGCHS distributions and a  copula density which is devised to encode between-square correlation among the variables involved in the sum. %\sout{densities functions of the sum of HS laws with the same or different kurtosis are worked out and a copula density is devised to encode between-square correlation among the variables involved in the sum.}
Section \ref{sec:app} and \ref{sec:forecast}  provide an empirical applications and Section \ref{sec:end} draws some conclusions. All proofs have been collected in an Appendix in order to ease the reading of the paper.  

% main text
\section{On the distribution of the sum of Gram-Charlier-like expansions of hyperbolic secant laws}\label{sec:theory}
In this Section, after recalling the main results of Gram-Charlier GCl expansions of an HS density (GCHS), we determine the density functions of the sum of $n$ of these independent densities (SGCHS hereafter). The intended result, when $n$ is even, can be also attained by summing $m$ independent convoluted linear hyperbolic (CLH) laws. SGCHS densities are worked out for both the cases of equal or different excess-kurtoses of the HS marginal laws which are added up. The effect introduced by the polynomial expansion in SGCHS densities is quantified and analyzed.
Finally, a copula density for SGCHS is devised to account for between-square dependence between the margins. 
The following Theorem covers the main results on the GC-like expansion of an HS law. 
\begin{thm}
The GC-like expansion of a standardized HS law
\begin{equation}\label{eq:hypsec}
f(x)=\frac{1}{2} \text{\text{sech}}\left(\frac{\pi x}{2}\right)
\end{equation}
accounting for an extra- kurtosis $\beta$ is given by 
\begin{equation}\label{eq:gchypsec}
\varphi (x,\beta )=\left(1+\frac{\beta }{\gamma _{4}} p_{4} (x)\right)f(x),
\end{equation}
where
%\begin{equation}\label{eq:poly}
%p_{4} (x)=x^{4} - ax^{2} +b
%\end{equation}
%is the fourth orthogonal polynomial associated to $f(x)$, with  parameters
%\begin{equation}
%a=\frac{m_{6} -m_{2} m_{4} }{m_{4} -m_{2}^{2} } =14   \qquad\qquad                
%b=\frac{m_{6} m_{2} -m_{4}^{2} }{m_{4} -m_{2}^{2} } =9
%\end{equation}
%depending on the moments  $m_j$  of  $f(x)$,  and
%\begin{equation}\label{eq:gamma4}
%\gamma _{4} =\int _{-\infty }^{\infty }p_{4}^{2}(x)f(x)dx=576
%\end{equation}
\begin{equation}\label{eq:poly}
p_{4} (x)= x^{4} -  14 x^{2} + 9 
\end{equation}
is the fourth orthogonal polynomial associated to $f(x)$, and $\gamma _{4} =\int _{-\infty }^{\infty }p_{4}^{2}(x)f(x)dx=576$.
The parameter $\beta$ measures the excess kurtosis of the variable $X$ with respect to the kurtosis of the parent HS law. The function $\varphi (x,\beta)$ is a density if $0 \leq \beta \leq 14.4$ and is unimodal as long as $\beta \leq 9,71$.
\end{thm}\label{thgclike}
\begin{proof}
See \cite{Bagnato2015a}.
\end{proof}
The kurtosis level attainable with the polynomial expansion of an HS law is much greater than what is obtainable for a Gaussian density. For the latter, the admissible boost in kurtosis cannot exceed 4 in order for the GC expansion to be a density and it must be lower than 2.4 to preserve unimodality (\citep{Bagnato2015a}).\\
As is well known, a portfolio is a set of several assets which often exhibit severe kurtosis. As such, those assets can be effectively modeled through the use of GCHS distributions, and the portfolio density can be modeled via the sum of its components. That is why in what follows we will derive the density of a sum of GCHS distributions. As a preliminary result, we will prove the following Theorem, which provides the density of the sum of $n$ independent HS distributions.
\begin{thm}\label{thmsum}
The density of the sum $Y=\sum _{i=1}^{n}X_{i}  $ of $n$ independent hyperbolic-secant variables $X_{i}\;\;\left(i=1,2,\dots,n\right)$ is
\begin{equation}\label{eq:sumhsodd}
g(y)=\frac{1}{2} \text{\text{sech}}\left(\frac{\pi }{2} y\right)\left[\frac{4^{m} }{(2m)!} \prod _{r=1}^{m}\left(\frac{y^{2} }{4} +\left(\frac{2r-1}{2} \right)^{2} \right) \right],
\end{equation}
if $n$ is odd, $n=2m+1$, and

\begin{equation}\label{eq:sumhseven}
g(y)=\frac{y}{2} \text{csch}\left(\frac{\pi }{2} y\right)\left[\frac{4^{m-1} }{(2m-1)!} \prod _{r=1}^{m-1}\left(\frac{y^{2} }{4} +r^{2} \right) \right],
\end{equation}
if $n$ is even, $n=2m$.
\end{thm}
\begin{proof}
See Appendix.
\end{proof}
In this regard, it is worth noting that the so-called convoluted linear hyperbolic (CLH) density function
\begin{equation}\label{eq:convoluted}
f(y)=\frac{y}{2} \text{csch}\left(\frac{\pi }{2} y\right),
\end{equation}
which arises from the convolution of two independent hyperbolic secant laws, is also the Fourier transform of the logistic function, and enjoys several desirable properties like bell shapedness, leptokurtosis and existence of moments and orthogonal polynomials of every order \cite{Bagnato2015a}. In particular, it is easy to prove the following

\begin{cor}\label{cor1}
The density of the sum of $m$ independent CLH distributions tallies with the density of the sum of $n=2m$ independent HS laws given in Equation \eqref{eq:sumhseven}.
\end{cor}
\begin{proof}
See Appendix.
\end{proof}
The kurtosis levels covered by \eqref{eq:sumhsodd} and \eqref{eq:sumhseven} can be broadened, to better match empirical data requirements, by duly modifying these laws via fourth-orthogonal polynomials, in the wake of Theorem \ref{thgclike}.\\
The coefficients of the polynomials at stake depend on the moments of the parent densities. They turn out to be cumbersome to compute and vary with the number of variables which are summed up. This can be overcome by moving from the GCl expansion of a sum of HS densities to the sum of GCHS densities (SGCHS). The following Theorem in fact establishes the density function of the sum of independent GCHS laws with the same excess kurtosis $\beta$. 

\begin{thm}\label{thmsgchsind}
Let $Y=\sum _{i=1}^{n}X_{i}$, where $X_{i} \;\;(i=1,2,{\dots},n) \sim \text{GCHS}(\beta)$ are assumed independent. The density function of $Y$ is given by

\begin{equation}\label{sgchsindodd}
g(y)=\frac{1}{2} \text{\text{sech}}\left(\frac{\pi }{2} y\right)\sum _{i=0}^{n}\sum _{j=i}^{n}\delta _{ij}   \frac{4^{(m+i+j)} }{[2(m+i+j)]!} \prod _{r=1}^{m+i+j}\left[\frac{y^{2} }{4} +\left(\frac{2r-1}{2} \right)^{2} \right],
\end{equation} 
if $n=2m+1$, and
\begin{equation}\label{sgchsindeven}
g(y)=\frac{y}{2} \text{csch}\left(\frac{\pi}{2} y\right)\sum _{i=0}^{n}\sum _{j=i}^{n}\delta _{ij}   \frac{4^{(m+i+j)-1} }{[2(m+i+j)-1]!} \prod _{r=1}^{m+i+j-1}\left[\frac{y^{2} }{4} +r^{2} \right],
\end{equation}
if $n=2m$. In both formulas, $\delta _{ij}$ is specified as
\begin{equation}
\delta _{ij} =(-2)^{j-i} {{j} \choose {i}}\sum _{k=j}^{n}{{n} \choose{k}} {{k} \choose {j}} \tilde{\beta}^{k},
\end{equation} 
with $\tilde{\beta}=\frac{\beta}{24}$.
\end{thm}

\begin{proof}
See Appendix.
\end{proof}

\begin{cor}\label{corsgchsind}
Alternative expressions for the density of the sum of GC-like expansions of $n$ hyperbolic-secant laws with the same excess kurtosis, specified as in Theorem \ref{thmsgchsind}, are
\begin{equation}
\label{eq12}
g(y)=\frac{1}{2} \text{\text{sech}}\left(\frac{\pi \; y}{2} \right)\sum _{j=0}^{2n}\theta _{j}  \frac{4^{(m+j-1)} }{(m+j)!} \prod _{r=1}^{m+j}\left[\frac{y^{2} }{4} +\left(\frac{2r-1}{2} \right)^{2} \right],
\end{equation} 
if $n=2m+1$, and
\begin{equation}
\label{eq13}
g(y)=\frac{y}{2} \text{csch}\left(\frac{\pi \; y}{2} \right)\sum _{j=0}^{2n}\theta _{j}  \frac{4^{m+j-1} }{(2m+j-1)!} \prod _{r=1}^{m+j-1}\left[\frac{y^{2} }{4} +r^{2} \right],
\end{equation} 
if $n=2m$. In both formulas, $\theta _{j}$ is specified as
\begin{equation}
\theta _{j} =(-1)^{j} \sum _{k=<j/2>}^{n} {{n}\choose{k}} {{2k}\choose{j}}\tilde{\beta}^{k},
\end{equation} 
with $\tilde{\beta}=\frac{\beta}{24} $ and $<j/2>$ denoting the smallest integer greater than, or equal to $j/2$. 
\end{cor}

\begin{proof}
See Appendix.
\end{proof}

\begin{figure}[ht!]
\begin{subfigure}
{\includegraphics[width=.5\textwidth]{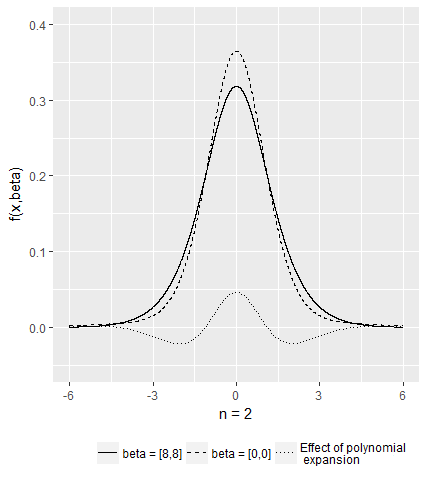}}
\end{subfigure}
\begin{subfigure}
{\includegraphics[width=.5\textwidth]{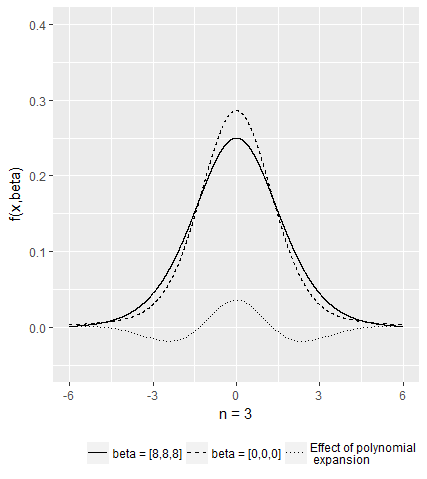}}
\end{subfigure}
\caption{Sum of two HS densities compared with the sum of the two corresponding GCHS with $\beta=8$, together with the effect of the polynomial expansion given in \eqref{effecteven} (left panel). Sum of three HS densities compared with the sum of the corresponding GCHS with $\beta=8$, together with the effect of the polynomial expansion given in \eqref{effectodd} (right panel).}
\label{fig:sgchs2}
\end{figure}

\noindent The graphs in Figure \ref{fig:sgchs1} depict the density functions of the sums of $n=2,3,4$ GC-like expansions of independent HS laws with the same excess kurtosis $\beta$.
\begin{figure}[ht!]
\centering
\includegraphics[width=\textwidth]{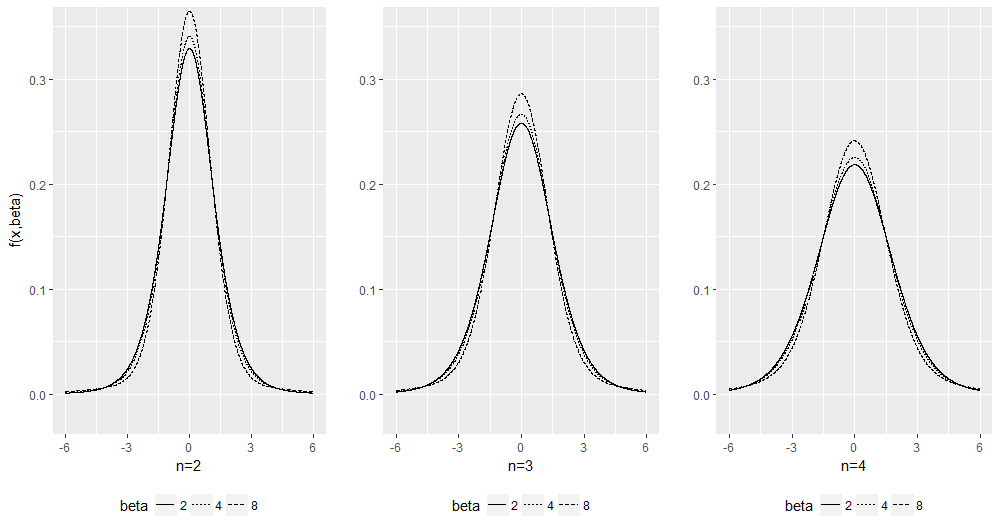}
\caption{SGCHS laws for $n = 2, 3, 4$ (left, central and right panel, respectively). The first panel shows the densities of the sums of three pairs of GCHS laws with $\bm\beta$ equal to [2,2], [4,4] [8,8], respectively. The second panel shows the densities of the sums of three triplets composed by GCHS laws with $\bm\beta$ equal to [2,2,2], [4,4,4] and [8,8,8], respectively. The third panel shows the densities of the sums of three quadruplets involving GCHS laws with $\bm\beta$ equal to [2,2,2,2], [4,4,4,4] and [8,8,8,8].}
\label{fig:sgchs1}
\end{figure}
\\
By comparing Equation \eqref{eq:sumhsodd} to Equation \eqref{eq12} and Equation \eqref{eq:sumhseven} to Equation \eqref{eq13}, we see that the factors
\begin{align}
\label{effectodd}
&\sum _{j=0}^{2n}\theta _{j}\frac{4^{(m+j-1)}}{(m+j)!} \prod _{r=1}^{m+j}\left[\frac{y^{2} }{4} +\left(\frac{2r-1}{2} \right)^{2} \right] \qquad\qquad&\text{if}&\qquad n=2m+1,\\
\label{effecteven}
&\sum _{j=0}^{2n}\theta _{j}  \frac{4^{m+j}}{(2m+j-1)!} \prod _{r=1}^{m+j-1}\left[\frac{y^{2} }{4} +r^{2} \right], \qquad\qquad&
\text{if}&\qquad n=2m
\end{align}
account for the polynomial expansions. 
Figure \ref{fig:sgchs2} depicts the role played by the factors on the sum of two and three GCHS densities with the same excess kurtosis $\beta=8$.

In general, the components of a portfolio are not likely to exhibit the same (extra) kurtosis. The following Corollary provides, in the wake of Theorem \ref{thmsgchsind}, the density function of the sum of independent GCHS laws when the excess kurtosis is no longer the same. 

\begin{figure}[ht!]
\centering
\includegraphics[width=\textwidth]{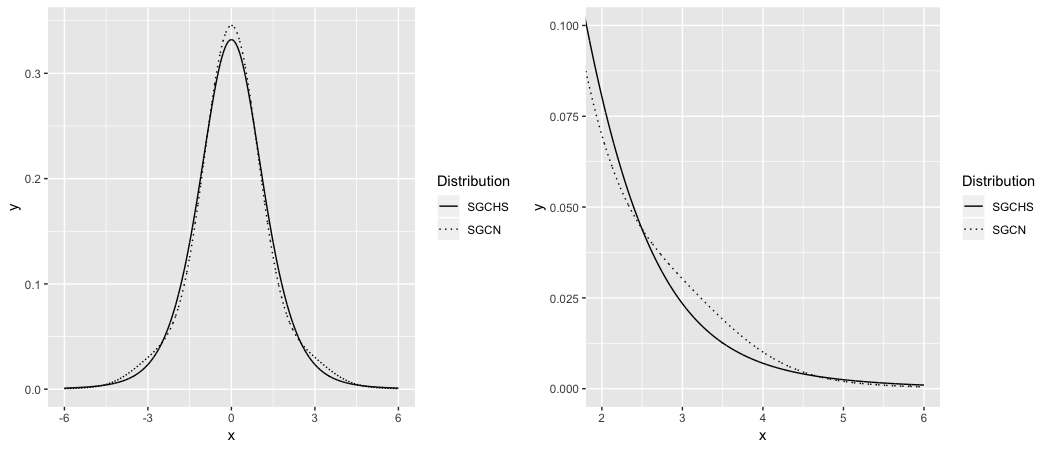}
\caption{SGCN and SCGHS obtained from the sum of two Gaussian and HS laws with the same excess kurtoses $\bm\beta=[2,4].$}
\label{fig:sgchs4}
\end{figure}

\begin{figure}[ht!]
\centering
\includegraphics[width=\textwidth]{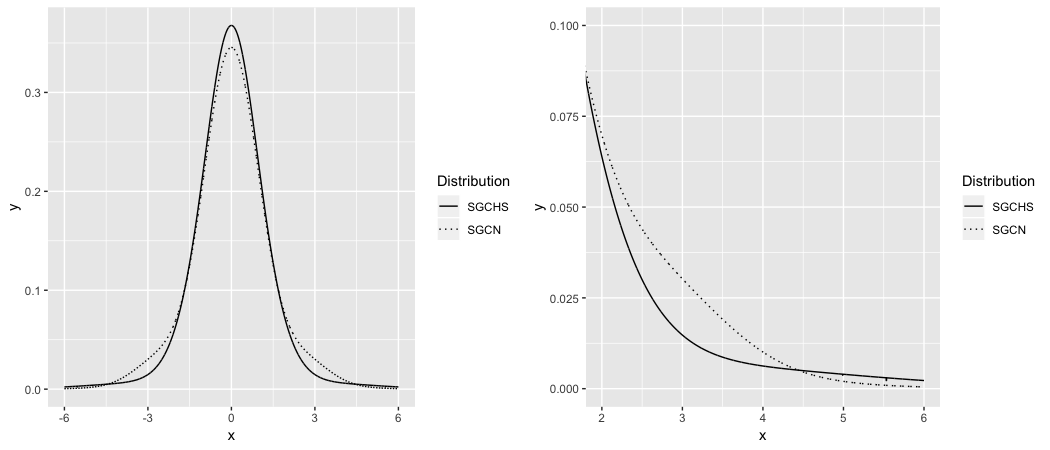}
\caption{SGCN and SCGHS obtained from the sum of three Gaussian and three HS laws with different excess kurtoses. The excess kurtoses are $\bm\beta=[2, 3, 3.5]$ for the SGCN and $\bm\beta=[8, 10, 11]$ for SGCHS. }
\label{fig:sgchs5}
\end{figure}

\begin{cor}\label{corsghs}
Let $Y=\sum _{i=1}^{n}X_{i}$, where $X_{i} \;\;(i=1,2,{\dots},n) \sim \text{GCHS}(\beta_i)$ are assumed independent. Then, the expressions of the density functions of $Y$  are those of \eqref{sgchsindodd}  or  \eqref{sgchsindeven}, when $n$ is odd or even, with  parameters $\delta _{ij}$ given by
\begin{equation}\label{deltasum}
\delta _{ij} =(-2)^{j-i} {{j}\choose {i}}\sum _{k=j}^{n}{{k} \choose {j}}b_{k},
\end{equation}
where
\begin{equation}\label{betakappa}
b_{k} =
\begin{cases} 1\qquad\qquad&\text{if}\qquad k=0 \\ 
\sum _{i_{1}=1}^{n}\tilde{\beta}_{i_{1}}\qquad\qquad&\text{if}\qquad k=1\\
\sum _{i_{1} =1}^{n-k+1}\sum _{i_{2=1} }^{i_{1} }\dots\sum _{i_{k=1} }^{i_{k-1} }\tilde{\beta}_{i_{1+k-1}}  \tilde{\beta }_{i_{2+k-2}} \dots \tilde{\beta }_{i_{k} }  \qquad\qquad&\text{if}\qquad  k=2,\dots,n,
\end{cases}
\end{equation}
with $\tilde{\beta}_{k} =\frac{\beta _{k}}{24}$.
\end{cor}
\begin{proof}
See Appendix.
\end{proof}
Figure \ref{fig:sgchs3} depicts SCGHS densities for different values of $n$ and different excess kurtoses $\beta_k$, and it provides evidence that the higher the excess kurtosis of the GCHS which are summed up, the more peaked and heavy tailed the distribution of the resulting sum. This effect dampens as the terms of the sum increase.
\begin{figure}[ht!]
\includegraphics[width=\textwidth]{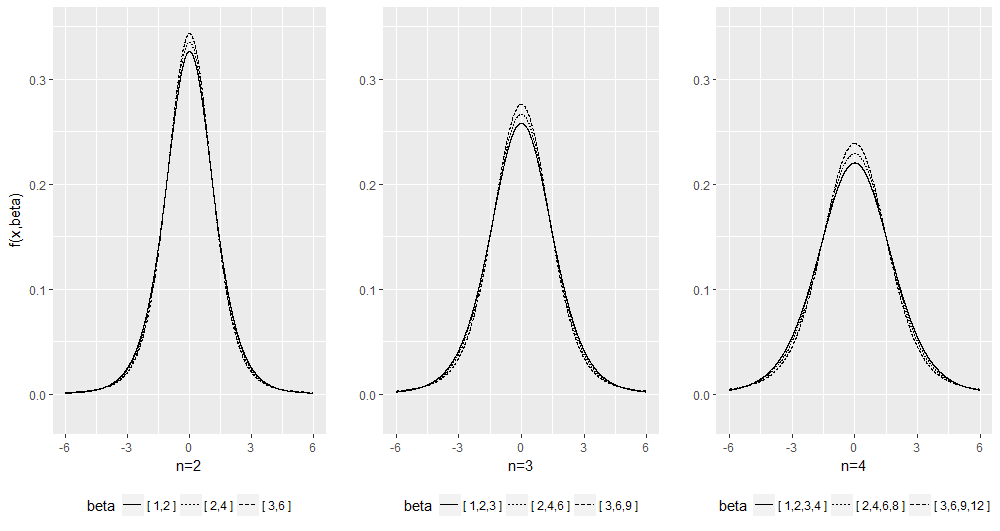}
\caption{SCGHS densities for $n = 2, 3, 4$ (left, central and right panel, respectively). In the first panel, the excess kurtosis of the two distributions involved in the sum are $\bm\beta=[1,2]$, $\bm\beta=[2,4]$ and $\bm\beta=[3,6]$, respectively. In the second panel the excess kurtoses of the three distributions entering the sum are $\bm\beta=[1,2,3]$, $\bm\beta=[2,4,6]$ and $\bm\beta=[3,6,9]$, respectively. In the third panel, four distributions with kurtoses equal to $\bm\beta=[1,2,3,4]$, $\bm\beta=[2,4,6,8]$ and $\bm\beta=[3,6,9,12]$ respectively, are summed up.}
\label{fig:sgchs3}
\end{figure}
The graph on the left of Figure \ref{fig:sgchs4} compares the sum of two GC expansions of Gaussian laws (SCGN hereafter) with the sum of two GC-like expansions of HS laws. The same excess kurtoses are assumed for the two distributions in each sum. The right panel focuses on the right tail of the distributions.
Similarly, the graph on the left of Figure \ref{fig:sgchs5} compares SGCN with SGCHS for $n=3$. Different excess kurtoses are assumed for the distributions which are summed up. The excess kurtoses of the three Gaussian laws are set equal to $\bm\beta=[2,3,3.5]$, while those of the three HS densities are set equal to $\bm\beta=[8,10,11]$. The right panel focuses on the right tail of the distributions.
\\
Looking at Figures \ref{fig:sgchs4} and \ref{fig:sgchs5}, we see that applying a polynomial expansion, inasmuch as it embodies the excess kurtosis of a distribution, forces a movement of probability mass from the shoulders of the parent distribution into the centre and tails of the resulting GC-like expansion \cite{Balanda1988} \cite{Finucan1964}. The higher the extra kurtosis, $\bm\beta$, or the more leptokurtic the parent law, the more substantial the probability displacement. For a given excess kurtosis this means, on the one hand, that the probability shift towards the peak and the tails is more accentuated in SGCHS than in SGCN laws and, on the other hand, that the probability loss in the shoulders is more substantial in the former than in the latter. Furthermore, this shift of probability becomes more significant as $\bm\beta$ increases. As we will see in Section \ref{sec:app}, this effect turns out to play an important role when the polynomially-modified distributions are targeted to evaluate risk measurements such as  the VaR or the ES.\\
Between-squares dependence also occurs when dealing with financial data. In order to take due account of this kind of dependence in the framework of SGCHS laws, we need to stretch beyond Theorem \ref{thmsgchsind} and devise a joint density function $\psi \left(x_{1} ,...,x_{n} \right)$ accounting for the intended dependence among its marginals $x_{1} ,...,x_{n} $. 
This can be done by introducing a suitable copula density, $c\left(u_{1} ,...,u_{n} \right)$ \cite{Nelsen2007}, in the density function of SGCHS. This copula is tailored to take into account between-squares correlations. The orthogonal polynomial technique, adopted to design GCl expansions of HS densities, paves the way to account for this kind of dependence, as shown in the following Theorem. 
\begin{thm}\label{sgchscopula}
Let $Y=\sum _{i=1}^{n}X_{i}$, where $X_{i} \;\;(i=1,2,{\dots},n) \sim \text{GCHS}(\beta_i)$. Then, the density function of $Y$, embodying between-squares dependence between two consecutive variables at a time, is as follows
\begin{equation}\label{copuladens}
g\left(y\right)=\int _{{\rm R}^{n-1} }\psi \left(\left(y-\sum _{i=2}^{n}x_{i}  \right),x_{2} ,...,x_{n} \right)dx_{2} ...dx_{n}
\end{equation}
where
\begin{equation}\label{copula}
\psi (x_{1} ,...,x_{n})=c\left(\Phi _{1} \left(x_{1} \right),...,\Phi _{n} \left(x_{n} \right);\beta _{1} ,...,\beta _{n} ,\gamma _{1} ,...\gamma _{n-1} \right)\prod _{i=1}^{n}\phi _{i} (x_{i}  ,\beta _{i})
\end{equation} 
and $\phi _{i} (x_{i} ,\beta _{i})$ are the GCHS densities as defined in Equation \eqref{eq:gchypsec}. The copula density $c(\cdot)$ is specified as 
\begin{equation}\label{copula2}
c\left(\Phi _{1} \left(x_{1} \right),...,\Phi _{n} \left(x_{n} \right);\beta _{1} ,...,\beta _{n} ,\gamma _{1} ,...\gamma _{n-1} \right)=
\prod _{i=1}^{n-1}\left[1+\gamma _{i} r_{i} \left(x_{i} ,\beta _{i} \right)r_{i+1} \left(x_{i+1} ,\beta _{i+1} \right)\right] 
\end{equation} 
with
\begin{equation}
\Phi _{i} \left(x\right)=\int _{-\infty }^{x}\phi (t,\beta _{i})dt
\end{equation} 
\begin{equation}
\label{eqr}
r_{i} (x,\beta _{i})=\frac{x^{2} -1}{4+\frac{\beta _{i} }{144} p_{4} (x)}.
\end{equation} 
Here,
\begin{equation}
\gamma _{i} =\mathbb E\left(X_{i}^{2} X_{i+1}^{2} \right)-1
\end{equation} 
denotes the between-squares correlation of the variates $X_{i}$ and $X_{i+1}$ subject to the following constraints \\
\begin{equation}\label{constraints}
-{1\mathord{\left/ {\vphantom {1 \max \left\{r_{i}^{-} r_{i+1}^{-} ,r_{i}^{+} r_{i+1}^{+} \right\}}} \right. \kern-\nulldelimiterspace} \max \left\{r_{i}^{-} r_{i+1}^{-} ,r_{i}^{+} r_{i+1}^{+} \right\}} \le \gamma _{i} \le -{1\mathord{\left/ {\vphantom {1 \min \left\{r_{i}^{-} r_{i+1}^{+} ,r_{i}^{+} r_{i+1}^{-} \right\}}} \right. \kern-\nulldelimiterspace} \min \left\{r_{i}^{-} r_{i+1}^{+} ,r_{i}^{+} r_{i+1}^{-} \right\}} 
\end{equation} 
\textit{where}
\begin{equation}r_{i}^{-} ={\mathop{\inf }\limits_{x}} r_{i} \left(x\right)   
\end{equation} 
and
\begin{equation}r_{i}^{+} ={\mathop{\sup }\limits_{x}} r_{i} \left(x\right).     
\end{equation} 
\end{thm}
\begin{proof}
See Appendix.
\end{proof}

Figure \ref{fig:nocciolo} shows the set of feasible values of $\gamma$ and $\bm\beta=[\beta_1,\beta_2]$ assuring  positiveness of the sum of two GCHS and GCN with the same excess kurtosis. Looking at the graph, we see that the feasible set of $\gamma$ and $\bm\beta$ assuring the positiveness of SGCHS is much wider than that of SGCN. 
\
\begin{figure}[ht!]
\centering
\includegraphics[width=\textwidth]{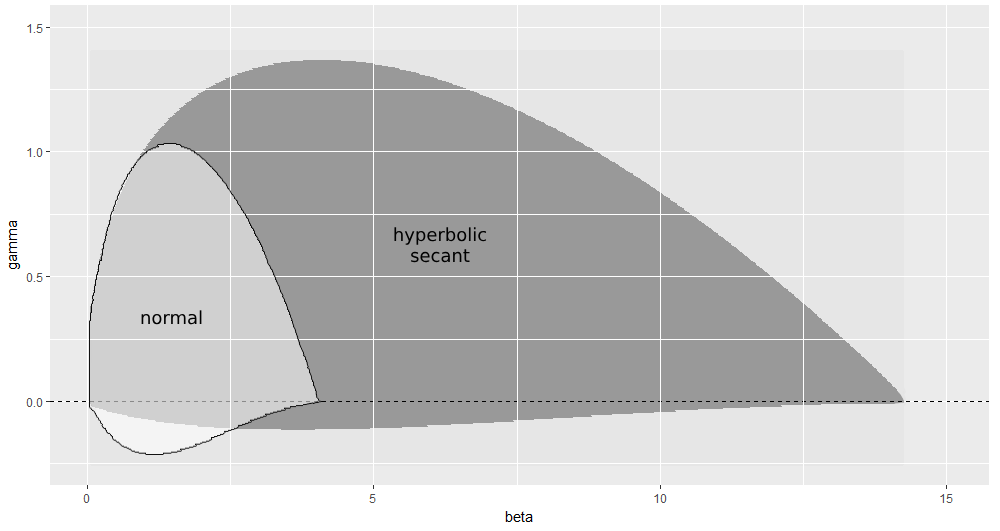}
\caption{Set of values for $(\bm\beta,\gamma)$ assuring the positiveness of the density of the sum of two GCHS laws (grey area) and the sum of two GCN laws (lighter grey area).}
\label{fig:nocciolo}
\end{figure}

\section{An empirical application}\label{sec:app}

The performance of SGCHS in modeling financial series and computing risk measures has been evaluated in two applications involving three assets: MSFT (Microsoft Corporation stock), \^{}N225 (Nikkei index) and NEM (Newmont Mining Corporation stock).
The observation period goes from 01/01/2011 to 31/12/2016 (T = 1420 days, excluding missing cases). Returns for the financial series at hand have been computed as  $r_{t} =\log \left(\frac{P_{t} }{P_{t-1} } \right)$, with $P_{t} $ denoting the closing price of each financial series on day $t$.\\
Table \ref{tab:desc} shows the descriptive statistics for the returns at stake.
\begin{table}
\centering\resizebox{0.5\textwidth}{!}{
\begin{tabular}{ccccc} \hline 
\textbf{Series} & \textbf{M} & \textbf{SD} & \textbf{Sk} & \textbf{K} \\ \hline 
MSFT & 0.00068 & 0.015 & -0.147 & 10.592 \\ 
\^{}N225 & 0.00043 & 0.015 & -0.551 & 8.306 \\ 
NEM & 0.00032 & 0.025 & -0.133 & 4.873 \\ \hline 
\end{tabular}}
\caption{Mean (M), standard errors (SD), skewness (Sk) and kurtosis (K) of three series MSFT, \^{}N225 and NEM.}
\label{tab:desc}
\end{table}\\
At first we considered a portfolio composed of two series [MSFT, \^{}N225], which were modeled by using four different distributions:
\begin{itemize}
\item two SGCHS laws, of which one with copula density and  the other without copula density (SGCHS-C and SGCHS, respectively); 
\item two SCGN densities, of which one with copula density and  the other without copula density (SGCN-C and SGCN, respectively).
\end{itemize}
The SGCN expansion of $n$ Gaussian laws either takes the following form \cite{Zoia2018}
\begin{equation}
\label{eq28}
f(y)=\sum _{j=0}^{n} {{n} \choose {j}} \left(\frac{\beta}{4!} \right)^{j} \frac{1}{\sqrt{2n\pi } } 
\left(\frac{1}{\sqrt{n} } \right)^{4j} e^{-\frac{y^{2} }{2n} } p_{4j} \left(\frac{y}{\sqrt{n} } \right),
\end{equation} 
if the excess kurtosis, $\beta$, is the same for all the GCN expansions, or the form
\begin{equation}
\label{eq29}
f(y)=\sum _{j=0}^{n}{{n} \choose {j}} \left(\frac{b_{n,j} }{{\rm (}4!{\rm )}^{j} } \right)\frac{1}{\sqrt{2n\pi } }
 \left(\frac{1}{\sqrt{n} } \right)^{4j} e^{-\frac{y^{2} }{2n} } p_{4j} \left(\frac{y}{\sqrt{n} } \right),
\end{equation} 
if the GCN expansions exhibit different excess kurtosis $\beta _{1},\beta _{2} ,\dots\beta _{n}$. The parameters $b_{n,j} $ in \eqref{eq29} are specified as follows
\begin{equation}
b_{n,j} =
\begin{cases} 1\qquad\qquad&\text{if}\qquad k=0 \\ 
\sum _{i_{1}=1}^{n}{\beta}_{i_{1}}\qquad\qquad&\text{if}\qquad k=1\\
\sum _{i_{1} =1}^{n-k+1}\sum _{i_{2=1} }^{i_{1} }\dots\sum _{i_{k=1} }^{i_{k-1} }{\beta}_{i_{1+k-1}}  {\beta }_{i_{2+k-2}} \dots{\beta }_{i_{k} }  \qquad\qquad&\text{if}\qquad k=2,\dots,n,
\end{cases}
\end{equation} 
In both \eqref{eq28} and \eqref{eq29} $p_{4j}$ denotes the $4j$-th degree Hermite polynomial 
\begin{equation}
p_{4j}(z)=z^{4j} +\sum _{i=1}^{2j}(-1)(2i-1)!! {{4j} \choose {2i}} z^{4j-2i}.
\end{equation} 

The sum of $n$ GCN laws, either with the same or with different kurtosis, which accounts for  between-squares dependence among variable as specified in Theorem \ref{sgchscopula}, is expressed in terms of the of GC expansions of Gaussian laws, namely

\begin{equation}
f(x)=\left(1+\frac{\beta}{4!} p_{4}(x)\right)\frac{1}{\sqrt{2\pi}} e^{-\frac{x^{2}}{2} }, 
\end{equation} 
where $p_{4}(x)=x^{4} -6x^{2} +3$. Furthermore, the term, $r_{i} (x,\beta _{i})$ takes, under normality, the form
\begin{equation}
r_{i} (x,\beta _{i})=\frac{x^{2} -1}{2+\frac{\beta _{i} }{12} p_{4}(x)}.
\end{equation} 
As both the MSFT and  \^{}N225 series  exhibit considerably high kurtoses, SCGHS laws are expected to provide a better fit for portfolio analysis.\\
In a further application, we indeed modeled the distribution of the portfolio covering all the three series [MSFT, \^{}N225, NEM], by using different laws. More precisely, for the triplet we employed:
\begin{itemize}
\item the densities of the sums of three GC expansions of Gaussian laws, of which one with copula density and the other without copula density, denoted as SGC3N-C and SGC3N, respectively.
\item the density of the sum of two GC-like HS laws and a GC of a Gaussian density. The SGCHS laws were used to model the pair of returns [MSFT, \^{}N225] which exhibit  severe kurtoses, and a SGCN was employed for the NEM series which exhibits moderate kurtosis and, as such does not recommend the use of SGCHS laws for its modeling. The density of this sum was considered both with and without copula specification. These two densities will be, respectively, denoted as SGCHSN-C and SGCHSN.
\end{itemize}

\begin{figure}[ht!]
\centering
\includegraphics[width=\textwidth]{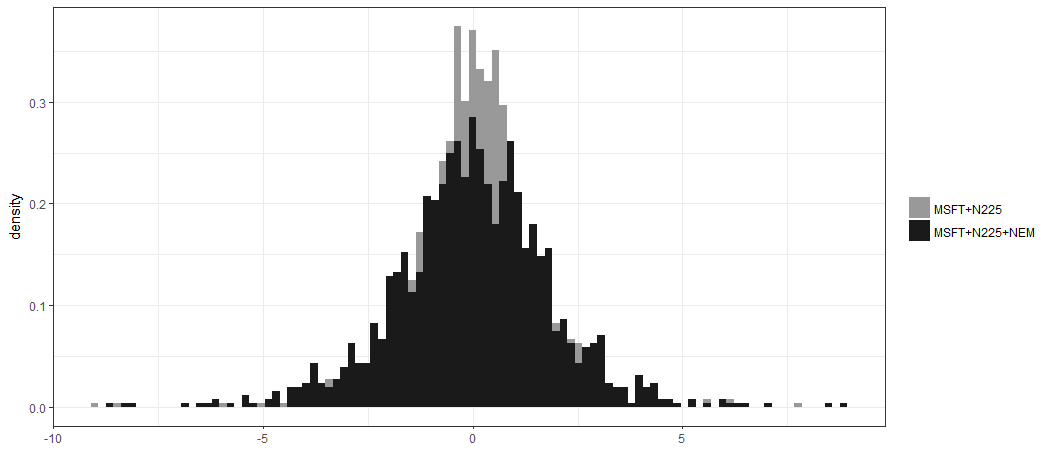}
\caption{Empirical distribution of the sum of the series [MSFT, \^{}N225] (grey) and of the three series MSFT, \^{}N225 and NEM (black).}
\label{fig:empirical}
\end{figure}

Figure \ref{fig:empirical} shows the empirical density histogram (or empirical density) of the sum of the pair [MSFT, \^{}N225] (grey) and the sum of the triplet [MSFT, \^{}N225, NEM] (black). It is clearly shown that adding a new series to the sum dampens the overall kurtosis.

The evaluation of the distributions adopted for the portfolio of the said returns has been used both on their goodness of fit to data and their performance in estimating common risk measures such as the Value at risk and the Expected shortfall.

\subsection{Modeling the sum of a pair of return series}\label{sec:pair}
\begin{center}
\begin{figure}[ht!]
\includegraphics[width=\textwidth]{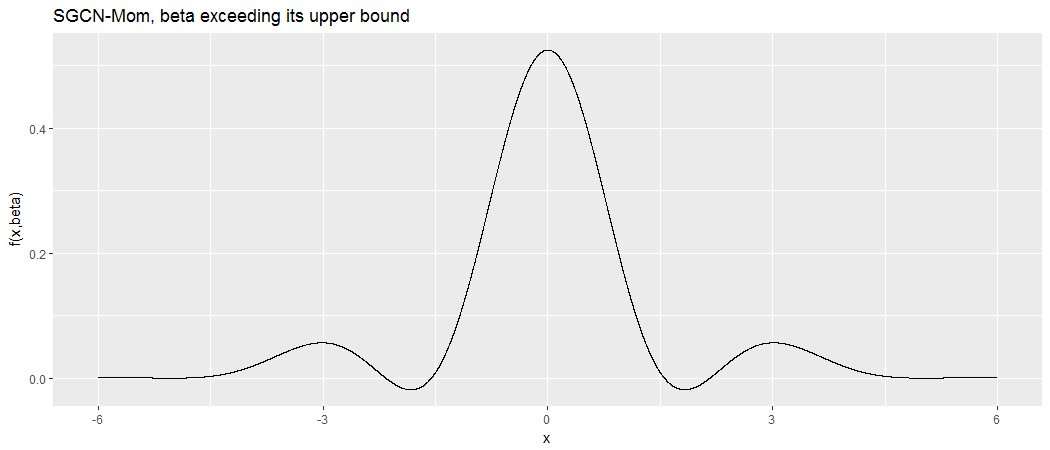}
\caption{$SGCN_{Mom}$ distribution for the sum [MSFT + \^{}N225].}
\label{fig:violation}
\end{figure}
\end{center}

\begin{figure}
{\includegraphics[width=1.05\textwidth]{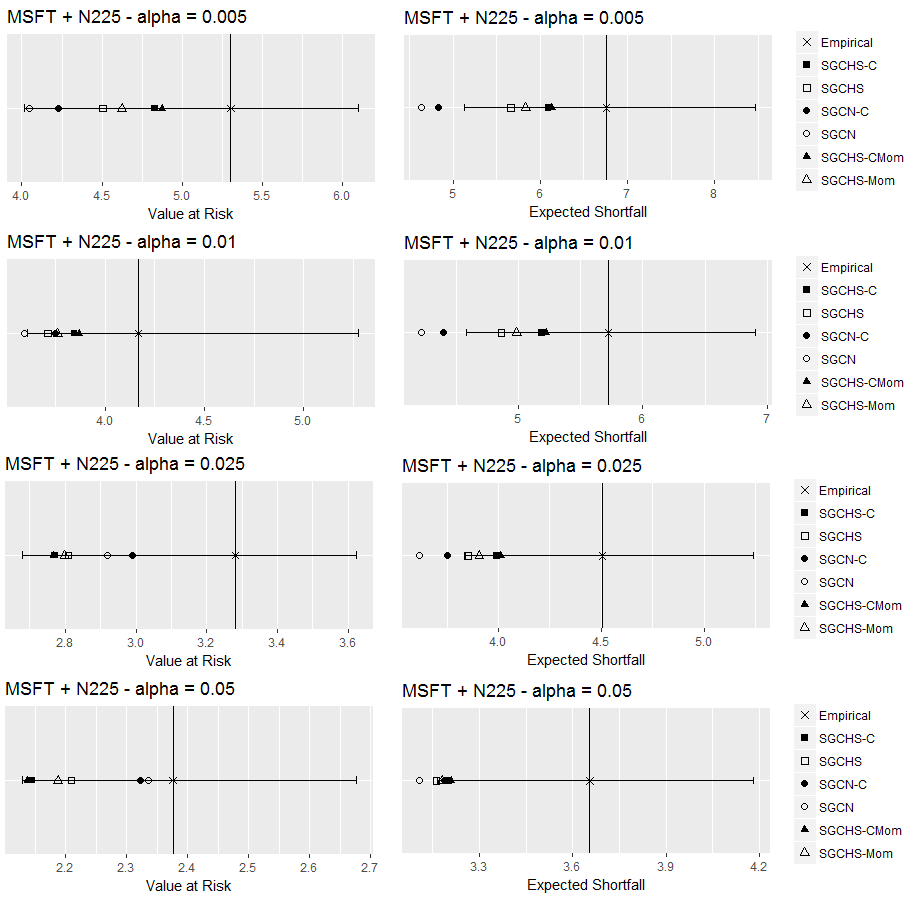}}
\caption{Empirical VaR and corresponding estimates (left column) and empirical ES and associated estimates (right column) obtained from SGCHS and SCGCN distributions fitted to the sum of the MSFT and \^{}N225 series.}
\label{fig:vares2}
\end{figure}

This section examines in detail the estimation process of the sum of the pair of returns [MSFT, \^{}N225] with the distributions specified at the beginning of Section \ref{sec:app}. To this aim, two estimation methods have been considered: the method of moments (Mom) and the maximum likelihood (ML) method. In the former case, the kurtosis parameters, $\bm\beta$, were estimated by using the empirical excess kurtoses with respect to the parent distributions (HS or Gaussian), while the $\gamma$ parameter was estimated by using its empirical analogous, namely
\begin{equation}
\widehat{\gamma}=\sum^T_{t=1}{\left(x^2_{it}x^2_{jt}\right)}-1
\end{equation} 
under the assumption to work with standardized series.
As for the latter estimation method, we have followed the IFM procedure. Following \cite{Joe1996}, the method is divided into two steps.
\begin{enumerate}
\item  First, we estimate the parameter of each marginal density $\phi _{i} (x_{i} ,\beta _{i} )$, maximizing the likelihood function
\begin{equation}
l_i({(x_{i,j})}_{j=1,\dots ,T};{\beta}_i)=\sum^T_{j=1}\mathrm{log} {\phi}_i(x_{i,j},\beta)\qquad i=1,\dots n\quad\quad
\end{equation} 
\item  Then, given the marginal estimates $({\hat{\beta }}_1,\dots ,{\hat{\beta }}_n)$, we estimate the copula parameters maximizing the likelihood function
\begin{align}
l_{c}& ((x_{i,j} )_{j=1,...,T} ;\hat{\beta }_{1} ,...,\hat{\beta }_{n} ,\gamma _{1} ,\dots,\gamma _{n-1})\\\nonumber
&=\sum_{j=1}^{T}\log c(\Phi _{1}(x_{1}),...,\Phi _{n} (x_{n});\hat{\beta }_{1} ,...,\hat{\beta }_{n} ,\gamma _{1} ,...,\gamma _{n-1})
\end{align} 
which, in our case, takes the form
\begin{align}
l_{c} &((x_{i,j})_{j=1,...,T} ;\hat{\beta }_{1} ,...,\hat{\beta }_{n} ,\gamma _{1} ,...,\gamma _{n-1})\\ \nonumber
&=\sum _{j=1}^{T}\log \prod _{i=1}^{n-1}\left[1+\gamma _{i} r_{i} (x_{i,j} ,\hat{\beta }_{i})r_{i+1} (x_{i+1,j} ,\hat{\beta }_{i+1} )\right] \\ \nonumber
&=\sum _{j=1}^{T}\sum _{i=1}^{n-1}\log\left[1+\gamma _{i} r_{i}(x_{i,j} ,\hat{\beta }_{i})r_{i+1}(x_{i+1,j} ,\hat{\beta }_{i+1})\right]\\ \nonumber
&=\sum _{i=1}^{n-1}\sum _{j=1}^{T}\log\left[1+\gamma _{i} r_{i}(x_{i,j} ,\hat{\beta }_{i})r_{i+1}(x_{i+1,j} ,\hat{\beta }_{i+1})\right]\\ \nonumber
&=\sum _{i=1}^{n-1}l_{c_{i}}((x_{i,j})_{j=1,...,T} ;\hat{\beta }_{1} ,...,\hat{\beta }_{n} ,\gamma _{i}).
\end{align} 
\end{enumerate}
This procedure allows the estimation of each single copula parameter at a time, for each couple of marginal variables in sequence.
The IFM estimator ${\widehat{\theta }}_n\ $ (obtained from a sample of size \textit{n}) of a vector $\bm\theta$, including all the unknown parameters, is asymptotically Gaussian, that is 
\begin{equation}
\sqrt{n}\left({\widehat{\bm\theta}}_n-\bm\theta \right){{\stackrel{d}{\rightarrow}}}N(0,G^{-1}\left(\bm\theta \right)),
\end{equation} 
where $G(\bm\theta)$ is the so-called Godambe information matrix which, following \cite{Joe1996}, is given by
\begin{equation}
\label{eqgam}
G\left(\bm\theta\right)=A^{-1}V(A^{-1})'.
\end{equation} 
Here, $A=\mathbb{E}\left[\frac{\partial s(\bm\theta)}{\partial \bm\theta}\right]$, $V=\mathbb{E}\left[s(\bm\theta )s(\bm\theta)'\right]$, and $s(\bm\theta)$ is the score function.\\ If analytical solutions are not available for Equation \eqref{eqgam}, the Godambe information matrix must be obtained via jackknife re-sampling. Common information criteria such as the AIC need to be adjusted accordingly \citep {Varin2011}, and the composite likelihood AIC (CLAIC) is employed to evaluate the goodness of fit.\\
Table \ref{tab:estpair} shows the estimation results for the four aforementioned distributions obtained with the two said estimation methods (Mom and ML). From the table we conclude that SGCHS distributions provide the best results in terms of CLAIC, especially when the copula parameter is included. Standard errors for the method of moments and empirical Godambe information matrices have been obtained via jackknife re sampling. As the excess kurtoses, $(\hat{\beta}_{1}= $, $\hat{\beta}_{2}= )$, of each series of the pair [MSFT, \^{}N225] exceeds the admissible boundary of the GCN distribution (see Figure \ref{fig:violation}), when calculated by using Mom, SGCN-Mom will not be taken into account in the analysis which follows.
\begin{center}
\begin{table}
\centering
\resizebox{\textwidth}{!}{
\begin{tabular}{cccccccc} \hline 
\textbf{Model} & \textbf{Parameter} & \textbf{Estimate} & \textbf{Std.Err.} & \textbf{z-value} & \textbf{$p$-value} & \textbf{CLAIC} & \textbf{LRT ($\gamma=0$)} \\ \hline 
$\text{SGCHS-C}_{\text{ML}}$ & ${\beta}_{HS1}$ & 4.027 & 0.997 & 4.635 & $<.001$ & 7583.7 & $<.001$ \\ 
 & ${\beta}_{HS2}$ & 3.354 & 1.020 & 3.289 & 0.001 &  &  \\  
 & ${\gamma}_{HS}$ & 0.710 & 0.162 & 4.394 & $<.001$ &  &  \\ \hline
$\text{SGCHS}_{\text{ML}}$ & ${\beta}_{HS1}$ & 4.027 & 0.964 & 4.177 & $<.001$ & 7609.9 & - \\ 
 & ${\beta}_{HS2}$  & 3.354 & 1.005 & 3.337 & $<.001$ &  &  \\ \hline
$\text{SGCN-C}_{\text{ML}}$ & ${\beta}_{N1}$ & 1.877 & 0.167 & 11.225 & $<.001$ & 7766.2 & $<.001$ \\ 
 & ${\beta}_{N2}$ & 1.509 & 0.166 & 9.075 & $<.001$ &  &  \\ 
 & ${\gamma}_{N}$ & 0.255 & 0.070 & 3.636 & $<.001$ &  &  \\ \hline
$\text{SGCN}_{\text{ML}}$ & ${\beta}_{N1}$ & 1.877 & 0.028 & 67.553 & $<.001$ & 7782.6 & - \\ 
 & ${\beta}_{N2}$ & 1.509 & 0.027 & 55.014 & $<.001$ &  &  \\ \hline
$\text{SGCHS-C}_{\text{Mom}}$ & ${\beta}_{HS1}$& 5.592 & 0.067 & 83.031 & $<.001$ & - & - \\ 
 & ${\beta}_{HS2}$& 3.306 & 0.051 & 64.425 & $<.001$ &  &  \\ 
 & ${\gamma}_{HS}$ & 0.539 & 0.006 & 84.498 & $<.001$ &  &  \\ \hline
$\text{SGCHS}_{\text{Mom}}$ & ${\beta}_{HS1}$ & 5.592 & 0.087 & 81.143 & $<.001$ & - & - \\ 
 & ${\beta}_{HS2}$ & 3.306 & 0.025 & 64.885 & $<.001$ &  &  \\ \hline
%$\text{SGCN}_{\text{Mom}}$ & ${\beta}_{N1}$& 7.592 & - & - & - & - & - \\ 
 %& ${\beta}_{N2}$& 5.306 & - & - & - & - & - \\ \hline 
\end{tabular}
}
\caption{Estimates of parameters $\bm\beta$ and $\gamma$ obtained with the maximum likelihood (ML) and moment (Mom) methods; Standard errors of the $\bm\beta$ estimates and $p$-values of the significance tests of their nullity; CLAIC and $p$-values of the likelihood ratio test (LRT) for the hypothesis $\gamma=0$.}
\label{tab:estpair}
\end{table}
\end{center}
To assess the goodness of fit to data of the aforementioned distributions, both the Kolmogorov-Smirnov (KS) and Anderson-Darling (AD) tests have been employed. The null hypothesis of both tests assumes that empirical and reference distributions are the same. As is well known, the AD test attributes more weight to the tails of a distribution than the KS test does. This is particularly relevant to our analysis as tails are the loci involved in the computation of risk measures. Looking at Tables \ref{tab:estpair} and \ref{tab:gofpair}, we conclude that both SGCHS and SGCHS-C perform better than the other distributions. 
\begin{table}
\centering
\resizebox{0.7\textwidth}{!}{
\begin{tabular}{ccccc} \hline 
\textbf{Model} & \textbf{AD} & \textbf{AD - $p$-value} & \textbf{KS} & \textbf{KS - $p$-value} \\ \hline 
$\text{SGCHS-C}_{\text{ML}}$ & 1.7000 & 0.1352 & 0.026 & 0.7790 \\
$\text{SGCHS}_{\text{ML}}$ & 1.3556 & 0.2151 & 0.026 & 0.7733 \\ 
$\text{SGCN-C}_{\text{ML}}$ & 1.3834 & 0.2069 & 0.027 & 0.7286 \\ 
$\text{SGCN}_{\text{ML}}$ & 2.1038 & 0.0806 & 0.033 & 0.4865 \\ 
$\text{SGCHS-C}_{\text{Mom}}$  & 1.7885 & 0.1204 & 0.026 & 0.7790 \\ 
$\text{SGCHS}_{\text{Mom}}$ & 1.4051 & 0.2009 & 0.026 & 0.7733 \\ \hline 
\end{tabular}
}
\caption{AD and KS statistics and $p$-values for the GC-like expansions of HS and Gaussian laws with and without copula density.}
\label{tab:gofpair}
\end{table}
Focusing now on tails: the performance has been evaluated of the aforementioned distributions in computing risk measures such as the VaR and the ES. The former provides the smallest value such that the probability of a (real-valued) random variable $X$ being at most this value is at least $1-\alpha $; that is 
\begin{equation}
VaR_{X}(\alpha)=inf\left\{x:F_{X} (x)\geq 1-\alpha \right\}=F_{X}^{-1} (1-\alpha)
\end{equation}
where $F_{X} (x)=P\left(X\le x\right)$ represents the cumulative distribution function of \textit{$X$}. \\
Unlike VaR, which is simply a threshold, ES provides information about the average loss beyond $VaR$ threshold. For a real-valued finite mean random variable $X$ with absolutely continuous cumulative distribution function, ES can be defined as follows \cite{Acerbi2002}:
\begin{equation}
ES_{X} (\alpha)=\mathbb E(X\left|X\ge VaR_{X} (\alpha)\right.)=\frac{\int _{VaR_{X} (\alpha )}^{+\infty }xdF_{X}(x) }{\int _{VaR_{X} (\alpha )}^{+\infty }dF_{X} (x) } =\frac{1}{\alpha} \int _{VaR_{X} (\alpha )}^{+\infty }xdF_{X}(x)
\end{equation} 
Figure \ref{fig:vares2} provides estimates of the Value at Risk, $\text{VaR}_{\alpha }$, computed by using the four mentioned distributions for $\alpha\in \{0.001, 0.005, 0.025, 0.05\}$. The empirical VaR, computed by using the empirical distribution of the sum MSFT+\^{}N225, $\text{VaR}_{\text{emp}}$, is depicted in the graphs with the upper, $\text{VaR}_{\text{emp}}^U$, and lower, $\text{VaR}_{\text{emp}}^L$, bounds of the bootstrap percentile intervals, at confidence $p=0.99$. The latter have been built by selecting $R = 1000$ block-bootstrap samples \citep{Kunsch1989} from the empirical distribution of the sum of the pair of returns [MSFT, \^{}N225]. The same has been done for the estimate, $\text{ES}_{\alpha}$, of the expected shortfall which has been compared to the empirical expected shortfall computed by using the empirical distribution, $\text{ES}_{\text{emp}}$. Bootstrap percentile intervals, $\text{ES}_{\text{emp}}^U$ and $\text{ES}_{\text{emp}}^L$, have been also computed for this risk measure following the procedure explained before. Detailed results of this procedure are given in a Table provided as supplementary material.
%\ref{tab:vares2} the Appendix.\\
Looking at Figure \ref{fig:vares2}, we conclude that SGCHS distributions prove effective in estimating VaR for $\alpha\in\{0.005, 0.01\}$, while SGCN perform better for $\alpha\in\{0.025, 0.05\}$. This can be explained by noticing that, according to Figures \ref{fig:sgchs4} and \ref{fig:sgchs5} in Section \ref{sec:theory}, quantiles which fall in the region pertaining to the shoulders of a distribution are better estimated by SGCN distributions since the latter are characterized by heavier shoulders than SGCHS laws. 
As far as ES estimates are concerned, SGCHS distributions are the only ones which always fall inside the confidence bands. In addition, we see that taking into account the between-square correlation via copula densities leads to better estimates of both these risk measures.

\subsection{Modeling the sum of a triplet of returns}
This Section examines in detail the problem of modeling the triplet [MSFT, N225, NEM], with the four distributions specified at the beginning of Section \ref{sec:app}. In this case, there are two possible copula parameters characterizing the polynomial expansion of the sum of the three densities: one to model the between-square dependence of the pair of series [MSFT, \^{}N225] and another to model the same type of dependence between \^{}N225 and NEM.\\
Table \ref{tab:esttriple} shows the results of the maximum likelihood estimation of the four distributions considered for the triplet, along with the estimates obtained using the method of moments (Mom), while Table \ref{tab:goftriple} provides the results of KS and AD tests. As before, $\text{SGC3N}_\text{Mom}$ will not be taken into account as the estimates of the excess kurtoses of the first two series exceed the boundary admissible for a Gaussian law.

\begin{figure}
{\includegraphics[width=1.05\textwidth]{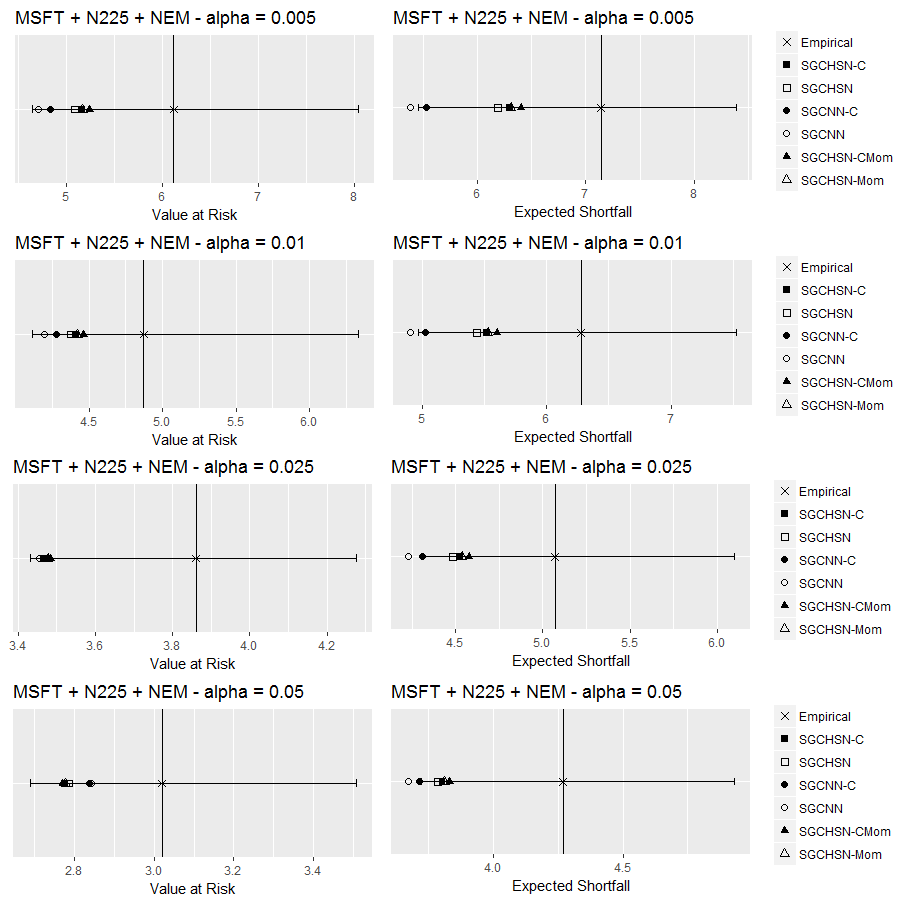}}
\caption{Empirical VaR and corresponding estimates (left column) and empirical ES and associated estimates (right column) obtained from SGCHSN and SCGCNN distributions fitted to the sum of the MSFT, \^{}N225 and NEM series.}
\label{fig:vares3}
\end{figure}

\begin{table}
\centering
\resizebox{\textwidth}{!}{
\begin{tabular}{cccccccc} \hline
\textbf{Model} & \textbf{Parameter} & \textbf{Estimate} & \textbf{Std.Err.} & \textbf{z-value} & \textbf{$p$-value} & \textbf{CLAIC} & \textbf{LRT ($\bm\gamma=\bm 0$)} \\ \hline 
  $\text{SGCHSN-C}_{\text{ML}}$ 
 & $\beta_{HS1}$ & 4.027 & 0.997 & 4.039 & $<.001$& 11499.0 & $<.001$\\ 
 & $\beta_{HS2}$ & 3.354 & 1.020 & 3.289 & $<.001$ &  &  \\ 
 & $\beta_{N3}$& 1.550 & 0.167 & 9.282 & $<.001$ &  &  \\ 
 & $\gamma_{1}$ & 0.710 & 0.162 & 4.381 & $<.001$ &  &  \\ 
 & $\gamma_{2}$ & 0.273 & 0.097 & 2.809 & 0.003 &  &  \\ \hline 
$\text{SGCHSN}_{\text{ML}}$ 
 & $\beta_{HS1}$ & 4.027 & 0.964 & 4.177 & $<.001$& 11534.7 &  \\ 
 & $\beta_{HS2}$ & 3.354 & 1.005 & 3.337 & $<.001$&  &  \\ 
 & $\beta_{N3}$ & 1.550 & 0.167 & 9.282 & $<.001$&  &  \\ \hline 
$\text{SGC3N-C}_{\text{ML}}$ 
 & $\beta_{N1}$ & 1.877 & 0.167 & 11.225 & $<.001$& 11695.2 & $<.001$\\
 & $\beta_{N2}$ & 1.509 & 0.166 & 9.075 & $<.001$&  &  \\ 
 & $\beta_{N3}$ & 1.550 & 0.167 & 9.282 & $<.001$&  &  \\  
 & $\gamma_{N1}$ & 0.255 & 0.070 & 3.636 & $<.001$&  &  \\ 
 & $\gamma_{N2}$ & 0.131 & 0.068 & 1.934 & $0.027$&  &  \\ \hline 
$\text{SGC3N}_{\text{ML}}$ 
 &$\beta_{N1}$ & 1.877 & 0.028 & 67.552 & $<.001$& 11706.4 &  \\ 
 & $\beta_{N2}$ & 1.509 & 0.027 & 55.014 & $<.001$&  &  \\ 
 & $\beta_{N3}$ & 1.550 & 0.167 & 9.282 & $<.001$&  &  \\\hline 
  $\text{SGCHSN-C}_{\text{Mom}}$ 
& $\beta_{HS1}$ & 5.592 & 0.067 & 83.031 & $<.001$& - & - \\
 & $\beta_{HS2}$ & 3.306 & 0.051 & 64.425 & $<.001$&  &  \\ 
 & $\beta_{N3}$ & 1.873 & 0.008 & 233.160 & $<.001$&  &  \\ 
 & $\gamma_{1}$ & 0.539 & 0.006 & 84.498 & $<.001$&  &  \\ 
 & $\gamma_{2}$ & 0.347 & 0.005 & 64.223 & $<.001$&  &  \\ \hline 
  $\text{SGCHSN}_{\text{Mom}}$ 
	& $\beta_{HS1}$ & 5.592 & 0.067 & 83.031 & $<.001$& - & - \\ 
 & $\beta_{HS2}$ & 3.306 & 0.051 & 64.425 & $<.001$&  &  \\ 
 & $\beta_{N3}$ & 1.873 & 0.008 & 233.160 & $<.001$&  &  \\ \hline 
\end{tabular}
}
\caption{Estimates of parameters $\bm\beta$ and $\gamma$ obtained with the maximum likelihood (ML) and moment (Mom) methods ; Standard errors of the $\bm\beta$ estimates and $p$-values of the significance tests of their nullity; CLAIC and $p$-values of the likelihood ratio test (LRT) for the hypothesis $\bm\gamma=\bm0$.}
\label{tab:esttriple}
\end{table}
\begin{table}
\centering
\resizebox{0.7\textwidth}{!}{
\begin{tabular}{ccccc}
\hline
\textbf{Model} & \textbf{AD} & \textbf{AD - $p$-value} & \textbf{KS} & \textbf{KS - $p$-value} \\ \hline 
$\text{SGCHSN-C}_{\text{ML}}$ & 1.1201 & 0.300 & 0.0213 & 0.931 \\ 
$\text{SGCHSN}_{\text{ML}}$ & 0.9825 & 0.367 & 0.0201 & 0.995 \\ 
$\text{SGC3N-C}_{\text{ML}}$ &  &  & 0.0211 & 0.9092 \\ 
$\text{SGC3N}_{\text{ML}}$ & 0.7690 & 0.504 & 0.0155 & 0.997 \\  
$\text{SGCHSN-C}_{\text{Mom}}$ & 1.3649 & 0.212 & 0.0240 & 0.851 \\ 
$\text{SGCHSN}_{\text{Mom}}$ & 1.1862 & 0.273 & 0.0223 & 0.905\\\hline
\end{tabular}
}
\caption{AD and KS statistics and $p$-values for the GC-like expansions of distributions assumed for the triplet [MSFT, \^{}N225, NEM].}
\label{tab:goftriple}
\end{table}
Looking at the Tables, we conclude that all densities have a similar behavior, with a slightly better performance for SGC3N-C. In general, including the copula significantly improvements CLAIC, over the distributions in which $\bm\gamma = \bm 0$.\\
Figure \ref{fig:vares3} provides estimates of VaR and ES, for $\alpha\in\{0.001, 0.005, 0.025, 0.05\}$ computed by using the four mentioned distributions together with the corresponding empirical estimates and the bootstrap percentile intervals. 
As before, SGCHSN laws perform better than other distributions in estimating both VaR and ES for every level of $\alpha$. The only exception is the VaR estimate for $\alpha=0.05$ where SGC3N laws, characterized by heavier shoulders (see Figures \ref{fig:sgchs4} and \ref{fig:sgchs5}, Section \ref{sec:theory}), show a better performance. As in the previous case, the estimates of both these risk measures, which are closer to their empirical counterparts, are obtained by taking into account the dependence among data via copula density.

\section{Out-of-sample performance of SGC distributions}\label{sec:forecast}
\begin{figure}
{\includegraphics[width=1.05\textwidth]{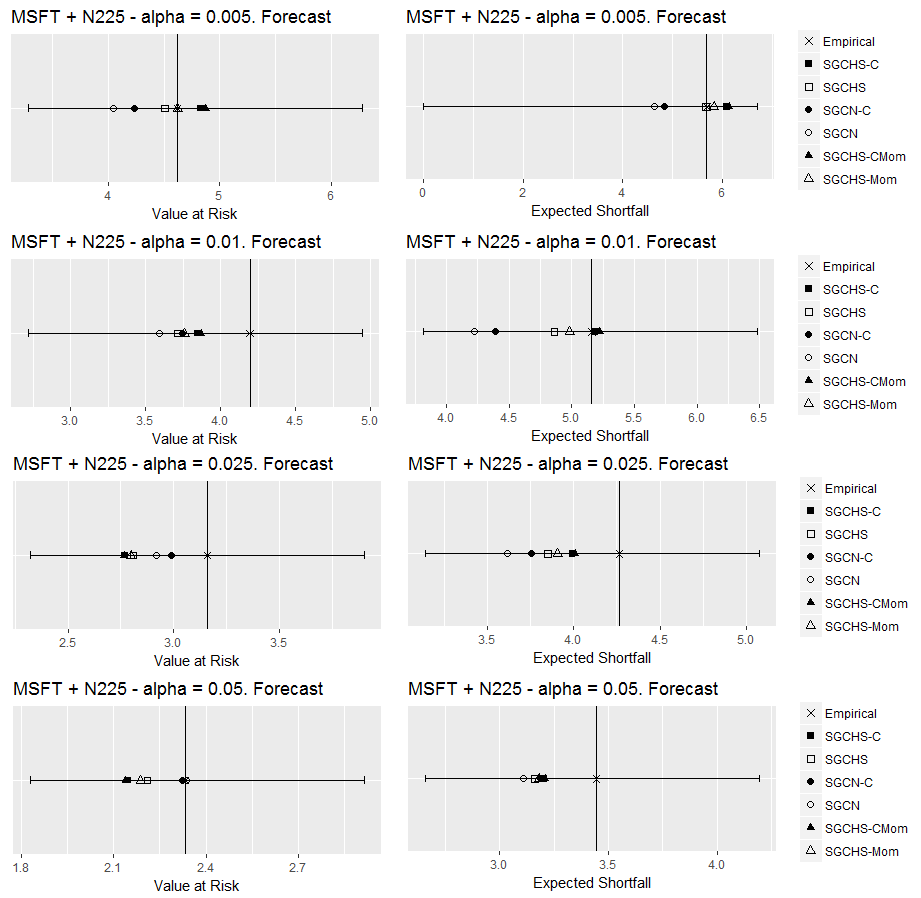}}
\caption{Out-of-sample empirical VaR and corresponding estimates (left column) and out-of-sample empirical ES and associated estimates (right column) obtained from SGCHS and SGCN distributions fitted to the sum of the MSFT and\^{}N225 series .}
\label{fig:fore2}
\end{figure}

\begin{figure}
{\includegraphics[width=1.05\textwidth]{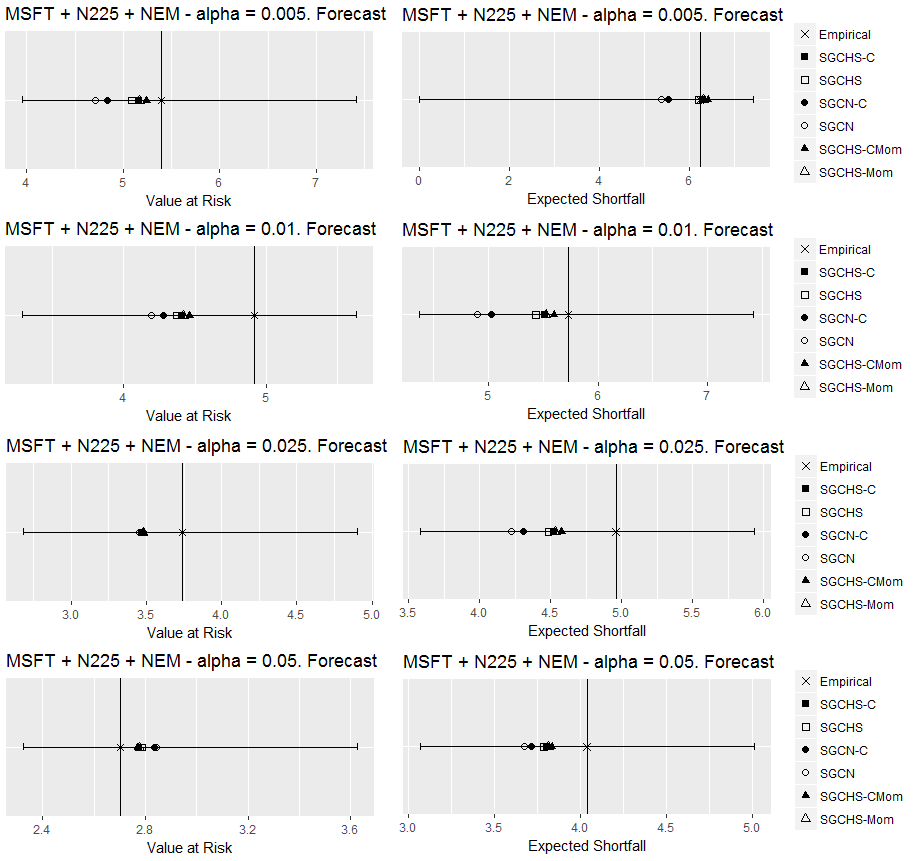}}
\caption{Out-of-sample empirical VaR and corresponding estimates (left column) and out-of-sample empirical ES and associated estimates (right column) obtained from SGCHSN and SGCNN distributions fitted to the sum of the MSFT, \^{}N225 and NEM series .}
\label{fig:fore3}
\end{figure}

In order to evaluate the out-of-sample performance of SGCHS distributions, we have computed the empirical quantiles of each financial series at hand, together with their bootstrap percentile intervals, by using data of series in a second time period, running from 01/01/2017 to 31/12/2018, not employed for the estimation of the SCGHS or SCGN distributions. In what follows we will call that part of the sample used to estimate the SCGHS or SCGN distributions first sample period and the other, used to assess the out-of-sample performance of the same, second sample period. Next, $\text{VaR}_\text{Emp}$ and $\text{ES}_\text{Emp}$ computed in the second sample period are compared to $\text{VaR}_\alpha$ and $\text{ES}_\alpha$ estimates obtained from the competing distributions in the first sample period. Figure \ref{fig:fore2} shows this comparison for the pair [MSFT, \^{}N225], while Figure \ref{fig:fore3} shows the results for the whole triplet [MSFT, \^{}N225, NEM].\\  
In case of the sum of the pair [MSFT, \^{}N225], the  VaR estimates obtained from SGCHS distributions are the most adequate for $\alpha \in\{0.005, 0.01\}$, whereas for $\alpha \in\{0.025, 0.05\}$ the estimates provided by SGCN distributions are more accurate. In terms of ES forecasting, SGCHS distributions are always the most accurate, although the gap with SGCN becomes narrower as $\alpha$ increases. For both these risk measures, the use of copula density leads to estimates which are closer to their empirical counterparts. When considering the sum of the triplet [MSFT, \^{}N225, NEM], VaR and ES estimates obtained by SGCHS densities always show the best performance. The copula parameter plays an important role in forecasting both VaR and ES for each level of $\alpha$.

The out-of-sample performance of SGC distributions in terms of VaR has been also evaluated via the Kupiec Coverage Test \citep{Kupiec1995}. The null hypothesis of the test assumes consistency between the percentage of losses that in the second sample period exceed $\text{VaR}_{\alpha}^{1}$, obtained from distributions estimated by using data from the first sample period, with the expected loss frequency for a given confidence level $\alpha$. The results of the test, given in Table \ref{tab:kupiec}, lead to the non-rejection of the null hypothesis for all SGCHS densities, while it is always rejected for SGCN and SGC3N lawswhen $\alpha = 0.005$ and between-square dependence is not taken into account.
\begin{center}
\begin{table}
\centering
\resizebox{\textwidth}{!}{
\begin{tabular}{cccccccccc} \hline 
\eject  &  & \multicolumn{2}{c}{$\alpha$=0.005} & \multicolumn{2}{c}{$\alpha$ =0.01} & \multicolumn{2}{c}{$\alpha$ =0.025} & \multicolumn{2}{c}{$\alpha$ =0.05} \\\hline
Series & Model & LR & $p$-value & LR & $p$-value & LR & $p$-value & LR & $p$-value \\ \hline 
MSFT + \^{}N225 
 & $\text{SGCHS-C}_{\text{ML}}$ & 0.0661 & 0.7971 & 0.9204 & 0.3374 & 0.3510 & 0.5535 & 3.3178 & 0.0685 \\
 & $\text{SGCHS}_{\text{ML}}$ & 0.1473 & 0.7012 & 1.8357 & 0.1755 & 0.3510 & 0.5535 & 2.0700 & 0.1502 \\
 & $\text{SGCN-C}_{\text{ML}}$ & 2.1868 & 0.1392 & 0.9204 & 0.3374 & 0.0966 & 0.7559 & 0.0575 & 0.8105 \\
 & $\text{SGCN}_{\text{ML}}$ & 3.8683 & \textbf{0.0492} & 3.0058 & 0.0830 & 0.0966 & 0.7559 & 0.0010 & 0.9749 \\
 & $\text{SGCHS-C}_{\text{Mom}}$ & 0.0661 & 0.7971 & 0.2962 & 0.5863 & 0.3510 & 0.5535 & 3.3178 & 0.0685 \\ 
 & $\text{SGCHS}_{\text{Mom}}$ & 0.1473 & 0.7012 & 0.9204 & 0.3374 & 0.3510 & 0.5535 & 2.0700 & 0.1502 \\ \hline 
MSFT + \^{}N225 + NEM 
 & $\text{SGCHSN-C}_{\text{ML}}$ & 0.1473 & 0.7012 & 3.0058 & 0.0830 & 0.3510 & 0.5535 & 0.0323 & 0.8575 \\ 
 & $\text{SGCHSN}_{\text{ML}}$ & 0.1473 & 0.7012 & 3.0058 & 0.0830 & 0.3510 & 0.5535 & 0.0323 & 0.8575 \\ 
 & $\text{SGC3N-C}_{\text{ML}}$ & 0.9122 & 0.3395 & 3.0058 & 0.0830 & 0.3510 & 0.5535 & 0.3728 & 0.5414 \\ 
 & $\text{SGC3N}_{\text{ML}}$ & 5.8890 & \textbf{0.0152} & 3.0058 & 0.0830 & 0.0966 & 0.7559 & 0.3728 & 0.5414 \\ 
 & $\text{SGCHS-C}_{\text{Mom}}$ & 0.1473 & 0.7012 & 3.0058 & 0.0830 & 0.3510 & 0.5535 & 0.0323 & 0.8575 \\ 
 & $\text{SGCHSN}_{\text{Mom}}$ & 0.1473 & 0.7012 & 3.0058 & 0.0830 & 0.3510 & 0.5535 & 0.0323 & 0.8575 \\ \hline 
\end{tabular}
}
\caption{Likelihood ratio test for SCG at different levels of $\alpha$.}
\label{tab:kupiec}
\end{table}
\end{center}
Furthermore, to assess the out-of-sample accuracy of VaR estimation via SGCHS densities, reference has been made to ABLF (average binary loss function) and AQLF (average quadratic loss function), which measure the number of observations of the second part of the sample, $x_t$ that exceed $VaR_{\alpha}^{1}$, defined as before, according to a specific loss function. The binary loss assigns a penalty of one for each exception, without considering its magnitude
\begin{equation}
BL =\begin{cases}
1 \quad&\text{if}\ \quad x_t\leq\text{VaR}(\alpha ). \\ 
0 &\text{otherwise} 
\end{cases}
\end{equation}
The quadratic loss function also considers the magnitude
\begin{equation}
QL =\begin{cases}
1+{\left(x_t-\text{VaR}(\alpha)\right)}^2 \quad&\text{if}\ \quad x_t\leq\text{VaR}(\alpha ). \\ 
0 &\text{otherwise} 
\end{cases}
\end{equation} 
Table \ref{tab:varf} provides estimates of the above loss functions for the pair [MSFT, \^{}N225] and the triplet [MSFT, \^{}N225, NEM]. Looking at the results we conclude that SGCHS distributions perform better for small values of $\alpha$, while SGCN and SGC3N are more competitive for higher levels of $\alpha$. 
Furthermore SGCHSN distributions offer the best overall performance for the triplet [MSFT, \^{}N225, NEM], in terms of both ABLF and AQLF, except for $\alpha=0.05$.
\begin{center}
\begin{table}
\centering
\resizebox{\textwidth}{!}{
\begin{tabular}{cccccccccc} \hline 
 &  & \multicolumn{2}{c}{$\alpha$ = 0.005} & \multicolumn{2}{c}{$\alpha$ = 0.01} & \multicolumn{2}{c}{$\alpha$ = 0.025} & \multicolumn{2}{c}{$\alpha$ = 0.05} \\ \hline 
Series & Model & ABLF & AQLF & ABLF & AQLF & ABLF & AQLF & ABLF & AQLF \\ \hline 
MSFT + \^{}N225 
 & $\text{SGCHS-C}_{\text{ML}}$ & \textbf{0.0042} & 0.0128 & \textbf{0.0147} & 0.0413 & 0.0294 & 0.1099 & 0.0692 & 0.2134 \\ 
 & $\text{SGCHS}_{\text{ML}}$ & 0.0063 & 0.0190 & 0.0168 & 0.0475 & 0.0294 & 0.1069 & 0.0650 & 0.2008 \\ 
 & $\text{SGCN-C}_{\text{ML}}$ & 0.0105 & 0.0279 & 0.0147 & 0.0443 & \textbf{0.0273} & \textbf{0.0923} & 0.0524 & 0.1748 \\  
 & $\text{SGCN}_{\text{ML}}$ & 0.0126 & 0.0340 & 0.0189 & 0.0537 & \textbf{0.0273} & 0.0969 & \textbf{0.0503} & \textbf{0.1711} \\
 & $\text{SGCHS-C}_{\text{Mom}}$ & \textbf{0.0042} & \textbf{0.0123} & 0.0147 & \textbf{0.0406} & 0.0294\textbf{} & 0.1102\textbf{} & 0.0692\textbf{} & 0.2144\textbf{} \\ 
 & ${SGCHS}_{Mom}$ & 0.0063 & 0.0174 & 0.0147 & 0.0438 & 0.0294 & 0.1077 & 0.0671 & 0.2057 \\ \hline 
MSFT +\^{}N225 +NEM 
 & $\text{SGCHSN-C}_{\text{ML}}$ & \textbf{0.0063} & 0.0179 & 0.0189 & 0.0465 & 0.0294 & 0.1066 & 0.0482 & 0.1946 \\ 
 & $\text{SGCHSN}_{\text{ML}}$ & \textbf{0.0063} & 0.0189 & 0.0189 & 0.0477 & 0.0294 & 0.1068 & 0.0482 & 0.1932 \\ 
 & $\text{SGC3N-C}_{\text{ML}}$ & 0.0084 &0.0225 & 0.0189 & 0.0477 & \textbf{0.0273} & \textbf{0.0947} & \textbf{0.0440} & \textbf{0.1755} \\ 
 & $\text{SGC3N}_{\text{ML}}$ & 0.0147 & 0.0322 & 0.0189 & 0.0459 & 0.0294 & 0.1045 & \textbf{0.0440} & 0.1805 \\ 
 & $\text{SGCHSN-C}_{\text{Mom}}$ &\textbf{0.0063} & \textbf{0.0169} & 0.0189 & \textbf{0.0448} & 0.0294 & 0.1054 & 0.0482 & 0.1954 \\ 
 & $\text{SGCHSN}_{\text{Mom}}$ & \textbf{0.0063} & 0.0178 & 0.0189 & 0.0461 & 0.0294 & 0.1059 & 0.0482& 0.1943\\ \hline 
\end{tabular}}
\caption{ABLF and AQLF indexes for the competing SGC distributions.}
\label{tab:varf}
\end{table}
\end{center}
Finally, the out-of-sample  $\text{ES}_\alpha$ estimates provided by SGC densities are assessed by implementing two tests: the McNeil and Frey test \cite{McNeil2000} and the Acerbi and Szekely test \cite{Acerbi2014}. The null hypotheses of both tests assume that the distribution used to evaluate ES tallies with the empirical one. This entails that, under the null hypothesis, the expected shortfall computed via SGCHS and SGCN distributions in the first part of the sample, $\text{ES}_{\alpha}^{1}$,  is a good estimate of the empirical ES computed with the data from the second period using $\text{VaR}_{\text{emp}}^{1}$.To perform the tests, block-bootstrap simulations have been implemented. In both cases, $B = 1000$ samples have been drawn from each empirical distribution. The statistics $Z_1$ and $Z_2$ are given by
%\noindent The statistic of the first test is
\begin{equation}
\label{eq:z1}
Z_{1} =\frac{1}{\sum _{t=1}^{N}I_{t}  } \frac{\sum _{t=1}^{N}X_{t} I_{t}  }{\text{ES}(\alpha)} +1
\end{equation} 
where 
\begin{equation}
I_{t} =
\begin{cases}
1 \quad\text{if}\quad X_t\leq\text{VaR}_{\text{emp}^{1}}\\ 
0 \quad\text{otherwise} \\
\end{cases}
\end{equation} 
\text{and}\\
%\noindent while the statistic of the other test is
\begin{equation}
\label{eq:z22}
Z_{2} =\frac{\sum _{t=1}^{T}X_{t} I_{t}}{N\alpha \text{ES}(\alpha)} +1,\quad t=1,2...,T
\end{equation} 
where $T$ denotes the sample size.
%\begin{equation}
%I_{t} =
%\begin{cases}
%1 \quad\text{if}\quad X_t\leq\text{VaR}_{\text{emp}^{1}} \\ 
%0 \quad\text{otherwise} 
%\end{cases}
%\end{equation} 
%To perform the tests, block-bootstrap simulations has been implemented. In both cases, $B = 1000$ samples have been drawn from the empirical distribution. The statistics $Z_1$ and $Z_2$ have been computed by using these samples and 
The bootstrapped $p$-value for the generic statistic $Z_j$, $j=1,2$ has been calculated as
\begin{equation}
p_{boot}=\frac{\sum _{b=1}^{B}I(Z_{j,b}^{*}<Z_{j}) }{B} 
\end{equation}
where $B$ is the number of bootstrap replications and $Z_{j,b}^{*}$ is the $b$-th bootstrap replicate and $Z_{j}$, $j=1,2$ is either the statistic \eqref{eq:z1} or the statistic \eqref{eq:z22} computed using the data of the second part of the sample.
Under $H_0$, the $Z_1$ and $Z_2$ are expected to be zero; hence, they signal a problem when they are negative and statistically significant.
Looking at Tables \ref{tab:esf1} and \ref{tab:esf2}, we conclude that SGCHS densities are the best models for both the MSFT + \^{}N225 pair  and the MSFT + \^{}N225 + NEM triplet: despite all distributions having both statistics not significantly lower than zero, SGCHS distributions almost always exhibit the lowest $Z_1$ and $Z_2$.
\begin{center}
\begin{table}
\centering
\resizebox{\textwidth}{!}{
\begin{tabular}{cccccccccc} \hline 
 &  & \multicolumn{2}{c}{$\alpha$=0.005} & \multicolumn{2}{c}{$\alpha$=0.01} & \multicolumn{2}{c}{$\alpha$=0.025} & \multicolumn{2}{c}{$\alpha$=0.05} \\ \hline 
Series & Model & $Z_1$ & $p$-value & $Z_1$ & $p$-value & $Z_1$ & $p$-value & $Z_1$ & $p$-value \\ \hline 
MSFT  + \^{}N225
 & $\text{SGCHS-C}_{\text{ML}}$ & -0.0090 & 0.4645 & \textbf{0.0052} & 0.4695 & -0.0920 & 0.5205 & -0.0904 & 0.5245 \\  
 & $\text{SGCHS}_{\text{ML}}$ & -0.0853 & 0.4645 & -0.0610 & 0.4855 & -0.1317 & 0.5205 & -0.1045 & 0.5245 \\  
 & $\text{SGCN-C}_{\text{ML}}$ & -0.2735 & 0.4645 & -0.1744 & 0.4855 & -0.1615 & 0.5205 & -0.0958 & 0.5245\\ 
 & $\text{SGCN}_{\text{ML}}$ & -0.3254 & 0.4655 & -0.2227 & 0.4855 & -0.2054 & 0.5205 & -0.1238 & 0.5245 \\ 
 & $\text{SGCHS-C}_{\text{Mom}}$ & \textbf{-0.0031} & 0.4695& 0.0124 & 0.4855 & \textbf{-0.0871} & 0.5205 & \textbf{-0.0882} & 0.5245\\ 
 & $\text{SGCHS}_{\text{Mom}}$ & -0.0539 & 0.4655 & -0.0350 & 0.4855 & -0.1157 & 0.5205 & -0.0982 & 0.5245 \\ \hline 
MSFT + \^{}N225 + NEM
 & $\text{SGCHSN-C}_{\text{ML}}$& -0.1781 & 0.3566 & -0.0135 & 0.5215 & -0.1188 & 0.5315 & -0.1628 & 0.5475 \\  
 & $\text{SGCHSN}_{\text{ML}}$& -0.1991 & 0.3566 & -0.0287 & 0.5215 & -0.1290 & 0.5315 & -0.1682 & 0.5475 \\
 & $\text{SGC3N-C}_{\text{ML}}$ & -0.2979 & 0.3566 & -0.0760 & 0.5215 & -0.1361 & 0.5315& -0.1529 & 0.5475 \\ 
 & $\text{SGC3N}_{\text{ML}}$ & -0.3492 & 0.3566 & -0.1188 & 0.5215 & -0.1783 & 0.5315 & -0.1888 & 0.5475 \\  
 & $\text{SGCHSN-C}_{\text{Mom}}$ & \textbf{-0.1587} & 0.3566 & \textbf{0.0016} & 0.5175 & \textbf{-0.1062} & 0.5315 & \textbf{-0.1543} & 0.5475 \\ 
 & $\text{SGCHSN}_{\text{Mom}}$ & -0.1760 & 0.3566 & -0.0117 & 0.5215 & -0.1163 & 0.5315 & -0.1600 & 0.5475 \\ \hline 
\end{tabular}}
\caption{The McNeil and Frey test assessing the out-of-sample performance of SGC densities in estimating ES (estimates and $p$-values).}
\label{tab:esf1}
\end{table}
\end{center}
\begin{center}
\begin{table}
\centering
\resizebox{\textwidth}{!}{
\begin{tabular}{cccccccccc} \hline 
 &  & \multicolumn{2}{c}{$\alpha$=0.005} & \multicolumn{2}{c}{$\alpha$=0.01} & \multicolumn{2}{c}{$\alpha$=0.025} & \multicolumn{2}{c}{$\alpha$=0.05} \\ \hline 
Series & Model & $Z_2$ & $p$-value & $Z_2$ & $p$-value & $Z_2$ & $p$-value & $Z_2$ & $p$-value \\ \hline 
MSFT + \^{}N225
 & $\text{SGCHS-C}_{\text{ML}}$ & 0.1539 & 0.4775 & -0.0427 & 0.5374 & -0.0073 & 0.5495 & -0.0516 & 0.5445 \\ 
 & $\text{SGCHS}_{\text{ML}}$ & 0.0899 & 0.4775 & -0.1122 & 0.5374 & -0.0439 & 0.5495 & -0.0651 & 0.5445\\  
 & $\text{SGCN-C}_{\text{ML}}$ & \textbf{-0.0680} & 0.4775 & -0.2310 & 0.5374 & -0.0714 & 0.5495 & -0.0567 & 0.5445 \\ 
 & $\text{SGCN}_{\text{ML}}$ & -0.1114 & 0.4775 & -0.2817 &0.5374& -0.1119 & 0.5495 & -0.0837 & 0.5445 \\ 
 & $\text{SGCHS-C}_{\text{Mom}}$ & 0.1589 & 0.4775 & \textbf{-0.0352} & 0.5374 & \textbf{-0.0028} & 0.5495 & \textbf{-0.0495} & 0.5445 \\  
 & $\text{SGCHS}_{\text{Mom}}$ & 0.1162 & 0.4775 & -0.0849 & 0.5374& -0.0291 & 0.5495 & -0.0591 & 0.5445 \\  \hline   
MSFT + \^{}N225 + NEM 
 & $\text{SGCHSN-C}_{\text{ML}}$ & 0.5060 & 0.3487 & -0.2748 & 0.5185 & -0.0321 & 0.5684 & 0.1224 & 0.5514 \\  
 & $\text{SGCHSN}_{\text{ML}}$ & 0.4972 & 0.3487 & -0.2939 & 0.5185 & -0.0414 & 0.5684 & 0.1183 &0.5514 \\ 
 & $\text{SGC3N-C}_{\text{ML}}$ & 0.4558 & 0.3487 & -0.3535 & 0.5185 & -0.0480 & 0.5684& 0.1299 & 0.5514 \\ 
 & $\text{SGC3N}_{\text{ML}}$ & \textbf{0.4343} & 0.3487 & -0.4073 & 0.5185 & -0.0869 & 0.5684 & \textbf{0.1028} & 0.5514 \\  
 & $\text{SGCHSN-C}_{\text{Mom}}$ & 0.5142 & 0.3487 & \textbf{-0.2558} & 0.5185 & \textbf{-0.0203} & 0.5684 & 0.1288 & 0.5514 \\ 
 & $\text{SGCHSN}_{\text{Mom}}$ & 0.5069 & 0.3487 & -0.2726 & 0.5185 & -0.0297 & 0.5684 & 0.1245 & 0.5514 \\ \hline 
\end{tabular}}
\caption{The Acerbi and Szekely test assessing the out-of-sample performance of SGC densities in estimating ES (estimates and $p$-values).}
\label{tab:esf2}
\end{table}
\end{center}
\section{Conclusion}\label{sec:end}
This paper introduces two significant improvements in the GC based approach to modeling portfolio distributions. First, it provides the distribution of a portfolio with risk factors distributed as GC-like expansions of the HS to account for possibly severe kurtosis. Second, it broadens the scope of this distribution via a novel copula, which is tailored to model between-square-dependence among risk factors. On the one hand,  moving from standard GC expansions of the Gaussian law to GCl expansions of the hyperbolic secant density, leads to novel distributions fit to capture the possibly severe kurtosis exhibited by several financial series. On the other hand, the introduction in the portfolio density of a copula, whose rationale still hinges on orthogonal polynomials, allows to account for non linear dependence among the returns of the portfolio, thus enabling portfolio distribution to fit in with the stylized factors of asset returns.
An empirical application of a portfolio composed of a set of international indexes provides evidence that the portfolio distributions here proposed are very tail sensitive densities which compare favorably with the  extant alternatives.
In particular, they provide more accurate estimates of both Value at risk and expected shortfall, especially for low confidence levels, than the standard GC-expansion based approach to portfolio modeling.  
\begin{appendix}
\section{}\label{sec:more}
The Fourier transform of an even function $f\left(x\right)$ is defined as follows
\begin{equation}\label{fourier}
F(\omega )=\int _{-\infty }^{+\infty }e^{i\omega x} f\left(x\right)dx =2\int _{0}^{+\infty }\cos \left(\omega x\right)f\left(x\right)dx =2F_{c}(\omega )
\end{equation}
where $F_{c} (\omega)$ denotes the so-called Fourier cosine transform, (see, e.g., \cite{Bateman1954}, Ch. 1). In functional notation, the Fourier cosine transform can be written as $F_{c} =T(f)$ and the inversion formula takes the form
\begin{equation}\label{inverseft}
T\left(F_c\right)=\frac{\pi}{2}f
\end{equation}
as otherwise stated 
\begin{equation}\label{invft2}
T\left(F\right)=\pi f
\end{equation} 
The following $f\left(x\right)\stackrel{T}{\longrightarrow}F_{c} \left(\omega \right)$ will denote that the functions on the right and left hand side form a pair of cosine Fourier transforms. 
\\Fourier transforms play an important role in statistics as they tally with the notion of characteristic functions of random variables while their anti-transforms are the densities functions of the same. In particular, we will focus on the sum $Y=\sum _{i=1}^{n}X_{i}$ of $n$ i.i.d distributed random variables. As is well known, the following proves true for the density $g(y)$ and the characteristic function $G(\omega)$ of the variable ${Y}$
\begin{equation}\label{convolute}
g=f^{(n)}=f*f*\dots*f \Leftrightarrow G=F^n
\end{equation}
where $f$ and $F$ are the density and characteristic function of each variable $X_{i},\;\; i=1,2,...,n$,  respectively, and $*$ denotes the convolution operation. We will now prove the statement of Theorem \ref{thmsum}, which concerns the density of the sum of $n$ i.i.d. hyperbolic secant laws.\\

\textit{Proof of Theorem \ref{thmsum}}\\
Let us begin by considering formula (1) on page 30 in \cite{Bateman1954}
\begin{equation}\label{bateman}
\text{sech}(ax)\stackrel{T}{\longrightarrow}\frac{1}{2} \pi \text{sech}\left(\frac{\pi }{2} a^{-1} \omega\right).
\end{equation}
By setting $a=1$ in \eqref{bateman} and taking into account \eqref{inverseft}, we conclude that 
\begin{equation}\label{fthypsec}
\frac{1}{2} \text{sech}\left(\frac{\pi }{2} x\right)\stackrel{T}{\longrightarrow}\frac{1}{2} \text{sech}(\omega).
\end{equation}
Equation \eqref{fthypsec}, in light of \eqref{fourier}, implies that $\text{sech}(\omega)$ is the density function of an hyperbolic secant law and, according to \eqref{convolute}, $\text{sech}(\omega)^n$ turns out to be the characteristic function of the sum of $n$ independent hyperbolic secant laws.\\ 
Now let us consider the following formulas on page 30 in \cite{Bateman1954}
\begin{align}\label{bateman2}
&{\left[\text{sech}(ax)\right]}^{2m+1}\stackrel{T}{\longrightarrow}\frac{2^{2m-1}\pi }{(2m)!a}\text{sech}\left(\frac{\pi \omega}{2a}\right)\prod^m_{r=1}{\left(\frac{\omega^2}{4a^2}+{\left(\frac{2r-1}{2}\right)}^2\right)} \\
&{\left[\text{sech}(ax)\right]}^{2m}\stackrel{T}{\longrightarrow}\frac{4^{2m-1}\pi \omega}{2\left(2m-1\right)!a^{2} }\text{csch}\left(\frac{\pi \omega}{2a}\right)\prod^{m-1}_{r=1}{\left(\frac{\omega^2}{4a^2}+r^2\right)}
\end{align}
By setting \textit{a}=1 in \eqref{bateman2} and taking into account \eqref{fourier} and \eqref{convolute}, the density function $g(y)$ of the sum for the case turns out to be as in Equation \eqref{eq:sumhsodd} when $n=2m+1$ and as in Equation  \eqref{eq:sumhseven} when $n=2m$. \qed\\

Theorem \ref{thmsum} paves the way to obtaining the density function of the sum of GC-like expansions of hyperbolic secant laws. This demands, as a preliminary result, the derivation of the characteristic function of a GC-like expansion, which is given in the following theorem.                       
\begin{thm}\label{chargc}
The characteristic function of a GC-like expansion specified as in \eqref{eq:gchypsec} is given by
\begin{equation}\label{tghgc}
F(\omega)=\text{sech}(\omega)\left\{1+\frac{\beta }{24} (\text{tgh}(\omega))^{4} \right\}
\end{equation}
\end{thm}
\begin{proof}
By setting $m=1$ and $a=1$ in \eqref{bateman2} and considering \eqref{inverseft}, we find that 
\begin{equation}
\frac{1}{2} \text{sech}\left(\frac{\pi }{2} x\right)\left(x^{2} +1\right) \stackrel{T}{\longrightarrow} (\text{sech}\mathrm{}{(\omega))}^3,
\end{equation}
which, bearing in mind \eqref{fthypsec}, leads to conclude that 
\begin{equation}\label{passage1}
\frac{1}{2}\text{sech}\left(\frac{\pi }{2} x\right)x^{2}\stackrel{T}{\longrightarrow}(\text{sech}(\omega))^{3} -\frac{1}{2} \text{sech}(\omega).
\end{equation}
Again, setting $m=2$ and $a=1$ in \eqref{bateman2} and taking into account \eqref{inverseft}, simple computations prove that     
\begin{equation}
\frac{2}{3} \text{sech}\left(\frac{\pi }{2} x\right)\left(\frac{x^{2} }{4} +\frac{1}{4} \right)\left(\frac{x^{2} }{4} +\frac{9}{4} \right)\stackrel{T}{\longrightarrow}(\text{sech}\mathrm{}{(\omega))}^5,
\end{equation}
which, bearing in mind \eqref{fthypsec} and \eqref{passage1}, yields 
\begin{equation}\label{passage2}
\frac{1}{2} \text{sech}\left(\frac{\pi }{2} x\right)x^{4} \stackrel{T}{\longrightarrow}\frac{1}{2}\left[24{(\text{sech}\mathrm{}(\omega))}^5-20{\left({\text{sech} \left(\omega\right)\ }\right)}^3+\text{sech}\mathrm{}(\omega)\right].
\end{equation}
In light of  \eqref{fthypsec}, \eqref{passage1}, \eqref{passage2}, and \eqref{eq:gchypsec}, the cosine transform of $\varphi(x,\beta)$ turns out to be
\begin{align}
\varphi (x,\beta )&=\left[1+\frac{\beta }{\gamma _{4} } (x^{4} -14x^{2} +9)\right]\frac{1}{2} \text{sech}\left(\frac{\pi x}{2} \right) \stackrel{T}{\longrightarrow}\\ \nonumber
&\frac{1}{2} \text{sech}(\omega)+\frac{\beta }{\gamma _{4} } \left\{12(\text{sech}(\omega))^{5} -10(\text{sech}(\omega))^{3} +\frac{1}{2} \text{sech}(\omega)\right\}+\\ \nonumber
&-\frac{14\beta }{\gamma _{4} } \left\{\left(\text{sech}(\omega)\right)^{3} -\frac{1}{2} \text{sech}(\omega)\right\}+\frac{9\beta }{2\gamma _{4} } \text{sech}(\omega),
\end{align}
and, taking into account that $\gamma _{4} =576$ , it can be worked out as follows
\begin{align}
&\frac{1}{2}\text{sech}(\omega)+\frac{\beta}{48}\left(\text{sech}(\omega)\right)^5-\frac{2\beta}{48}(\text{sech}(\omega))^3+\frac{\beta}{48}\text{sech}(\omega)\\ \nonumber
&=\frac{1}{2}\text{sech}(\omega)+\frac{\beta}{48}\text{sech}(\omega)[1-2(\text{sech}(\omega))^2+(\text{sech}(\omega))^4]\\ \nonumber
&=\frac{1}{2}\text{sech}(\omega)+\frac{\beta}{48}\text{sech}(\omega)\left\{1-(\text{sech}(\omega))^{2}\right\}^{2}\\ \nonumber
&=\frac{1}{2}\text{sech}(\omega)+\frac{\beta}{48}\text{sech}(\omega)[(\text{tanh}(\omega))^2]^2
\end{align}
This proves \eqref{tghgc}, in light of \eqref{fourier}.                                                                                      
\end{proof}
\textit{Proof of Corollary \ref{cor1}}\\
According to formula (2) on page 30 in \cite{Bateman1954}, the Fourier transform of the convoluted hyperbolic-secant law, $\frac{y}{2} \text{csch}\left(\frac{\pi}{2} y\right)$ is $(\text{sech}(\omega))^3$. This, taking into account \eqref{convolute},  entails that the characteristic function of the density of the sum of $m$ independent convoluted hyperbolic-secant laws is $(\text{sech}(\omega))^{3m}$. Hence, by using formula \eqref{bateman2} and bearing in mind \eqref{invft2} we obtain formula \eqref{eq:convoluted}\qed.
\\\\
Thanks to this result, we can now prove Theorem \ref{thmsgchsind} and Corollary \ref{corsgchsind}.\\

 \textit{Proof of Theorem \ref{thmsgchsind}.}\\
According to \eqref{convolute} and \eqref{tghgc}, the Fourier transform of the sum of $n$ independent hyperbolic secants is 
\begin{align}
[F(\omega)]^n
&={(\text{sech}(\omega))}^n\{1+\widetilde{\beta}[1-(\text{sech}(\omega))^2]^2\}^n \\ \nonumber
&={(\text{sech}(\omega))}^n\{1+\widetilde{\beta}[1-2(\text{sech}(\omega))^2+(\text{sech}(\omega))^4]\}^n,
\end{align}
where $\tilde{\beta}=\frac{\beta }{24}$.
The above formula, with some computations, can be worked out as follows
\begin{align}\label{triplesum}
[F(\omega)]^{n}
&=\text{sech}(\omega)^{n}\sum_{k=0}^{n}{{n} \choose {k}}\tilde{\beta}^{k}[1-2\text{sech}(\omega)^{2}+\text{sech}(\omega)^{4}]^{k} \\ \nonumber
&=\text{sech}(\omega)^{n}
  \sum_{k=0}^{n}{{n} \choose {k}}\tilde{\beta}^{k} \nonumber
  \sum_{j=0}^{k}{{k} \choose {j}}\text{sech}(\omega)^{2j} \nonumber
  \sum_{i=0}^{j}{{j} \choose {i}}\text{sech}(\omega)^{2i}(-2)^{j-i}.
\end{align}
Now, as it can be proved that (see \citep{Graham1994}, p.36)
\begin{equation}
\sum _{k=0}^{n}\sum _{j=0}^{k}\sum _{i=0}^{j}a_{ijk}=\sum _{i=0}^{n}\sum _{j=i}^{n}\sum _{k=j}^{n}a_{ijk},
\end{equation}
equation \eqref{triplesum} can be written as
\begin{equation}\label{deltas}
[F(\omega)]^{n} =\sum _{i=0}^{n}\sum _{j=i}^{n}\text{sech}{\rm (}\omega {\rm )}^{n+2\left(i+j\right)}  \delta _{ij}
\end{equation}
where
\begin{equation}\label{deltas2}
\delta _{ij} =(-2)^{j-i} {j \choose i}\sum_{k=j}^{n}{n \choose k}{j \choose j}\tilde{\beta}^{k}
\end{equation}
with $\tilde{\beta }=\frac{\beta}{24}$.
Given equation \eqref{deltas}, use of \eqref{bateman2} can be made to work out the density of the sum of $n$ independent GC-like expansions of linear hyperbolic laws. This leads to Equation \eqref{sgchsindeven} when $n=2m$ and to Equation \eqref{sgchsindodd} when $n=2m+1$, respectively.\qed\\

\textit{Proof of Corollary \ref{corsgchsind} }\\
Equation \eqref{triplesum} can be also worked out as 
\begin{align}\label{triplesum2}
[F(\omega)]^{n}&=\text{sech}(\omega)^{n} \sum _{k=0}^{n}{n \choose k} \tilde{\beta}^{k}[1-\text{sech}(\omega)^{2}]^{2k}\\ \nonumber
&=\text{sech}(\omega)^{n}\sum_{k=0}^{n}{n \choose k} \tilde{\beta}^{k} \sum _{j=0}^{2k}{k \choose j}(-1)^{j} \text{sech}(\omega)^{2j}.
\end{align}
In fact, as some computations show, the following identity holds true
\begin{equation}
\sum _{k=0}^{n}\sum_{j=0}^{2k}a_{k} b_{j}=\sum _{j=0}^{2n}b_{j}\sum _{k=<j/2>}^{n}a_{k}
\end{equation}
and formula \eqref{triplesum2} can also be written as follows
\begin{equation}
F(\omega )=\sum _{j=0}^{2n}\text{sech}(\omega )^{n+2j}\theta _{j}
\end{equation}
where
\begin{equation}
\theta _{j} =(-1)^{j}\sum_{k=<j/2>}^{n}{n \choose k} {{2k} \choose j}\tilde{\beta}^{k}
\end{equation}
with $<j/2>$ being the smallest integer greater or equal to $j/2$. Next, putting forward the same argument advanced in Theorem \ref{thmsgchsind}, we easily obtain the densities \eqref{sgchsindodd} and \eqref{sgchsindeven} of Corollary \ref{corsgchsind}.\qed\\  
\\
\textit{Proof of Corollary \ref{corsghs}}\\
According to Theorem \ref{chargc}, the Fourier transform of the sum of $n$ independent GC-like expansions with different excess kurtosis $\beta _{k},\;\;k=1,2,\dots,n$ is given by
\begin{align}\label{sgchsdiffbeta}
&\prod _{k=1}^{n}F_{k} (\omega) =\text{sech}(\omega)^{n} \prod _{k=1}^{n}\left\{1+\frac{\beta _{k} }{24} \text{tgh}(\omega)^{4} \right\} \\\nonumber
&=\text{sech}(\omega)^{n}\prod_{k=1}^{n}\left\{1+\tilde{\beta}_{k} \left[1-\text{sech}(\omega)^{2}\right]^{2} \right\} \\ \nonumber
&=\text{sech}(\omega)^{n}\sum _{k=0}^{n}b_{k}[1-2\text{sech}(\omega)^{2}+\text{sech}(\omega)^{4}]^{k}
\end{align}
where $\tilde{\beta}_{k} =\frac{\beta _{k}}{24} $ and  the $b_{k}$'s are defined as in \eqref{betakappa} (\citep{Nyblom1999}, p. 5). 
Then, with some computations, Equation \eqref{sgchsdiffbeta} can be worked out as follows
\begin{align}
\prod _{k=1}^{n}F_{k}(\omega)=\text{sech}(\omega)^{n}
\sum_{k=0}^{n}b_{k}\sum_{j=0}^{k}{{k} \choose {j}}\text{sech}(\omega)^{2j}
\sum_{i=0}^{j}{{j} \choose {i}}\text{sech}(\omega)^{2i}(-2)^{j-i}
\end{align}
and by applying Equation \eqref{deltas} it can be rewritten as in \eqref{deltas2} with $\delta _{ij}$'s specified as in \eqref{deltasum}. 
As far as the derivation of the density of the sum variable ${Y}$ is concerned, the proof follows the same argument as that of Theorem \ref{thmsgchsind}.\qed\\  

\textit{Proof of Theorem \ref{sgchscopula}}\\
The rationale of \eqref{copuladens} can be explained by noting that the density function of the random vector $(X_{1} ,...,X_{n})$ embodying between-square dependance among its margins $X_{1} ,...,X_{n}$ can be written as 
\begin{equation}\label{jointdens}
\psi (x_{1} ,\dots,x_{n})=\prod _{i=1}^{n}\phi (x_{i},\beta _{i} )\prod _{j=1}^{n-1}\left[1+\gamma_{i} r_{i} (x_{i})r_{i+1} (x_{i+1})\right]
\end{equation}
where 
\begin{equation}\label{constraint2}
\left[1+\gamma _{i} r_{i} (x_{i} )r_{i+1} (x_{i+1} )\right]
\end{equation}
In \eqref{constraint2}, $r_{i}(x_{i})r_{i+1}(x_{i+1})$ is defined as in \eqref{eqr} and $\gamma _{i}$ denotes the correlation coefficient of the (standardized) variates $X_{i}$ and $X_{i+1}$. Some computations prove that condition \eqref{constraints} is necessary for the positiveness of \eqref{jointdens}, which is mandatory in order for \eqref{copuladens} to represent a density function. Furthermore, upon noting that formula \eqref{constraints} can be written as 
\begin{equation}
\left[1+\gamma _{i} r_{i} (x_{i} )r_{i+1} (x_{i+1} )\right]=1+\gamma _{i} \frac{p_{2} (x_{i} )p_{2} (x_{i+1} )}{\left\| p_{2} \right\| \left(1+\frac{\beta }{\gamma _{4} } p_{4} (x_{i} )\right)\left(1+\frac{\beta }{\gamma _{4} } p_{4} (x_{i+1} )\right)}
\end{equation}
where $p_{2} (x)=x^{2} -1$ is the second-order orthogonal polynomial associated to a HS law and $\left\| p_{2} \right\| =
\int _{-\infty }^{\infty }p_{2} (x)^{2}  f(x)dx=16$ is its squared norm playing the role of normalization factor, the same argument put forward in Faliva et al (2016) can be used to prove that \eqref{copuladens} is a density function embodying both leptokurtosis in the margins and between-square correlation among its margins in due sequence. Hence, the density function of the linear transformation $Y=\sum _{i=1}^{n}X_{i}$ of the random vector $(X_{1} ,\dots,X_{n})$ can be obtained along the same lines as in  \cite{Gut2013} on page 10. 
\end{appendix}
\bibliographystyle{elsarticle-harv} 
\bibliography{SUM_GCHS}

\begin{thebibliography}{24}
\expandafter\ifx\csname natexlab\endcsname\relax\def\natexlab#1{#1}\fi
\providecommand{\url}[1]{\texttt{#1}}
\providecommand{\href}[2]{#2}
\providecommand{\path}[1]{#1}
\providecommand{\DOIprefix}{doi:}
\providecommand{\ArXivprefix}{arXiv:}
\providecommand{\URLprefix}{URL: }
\providecommand{\Pubmedprefix}{pmid:}
\providecommand{\doi}[1]{\href{http://dx.doi.org/#1}{\path{#1}}}
\providecommand{\Pubmed}[1]{\href{pmid:#1}{\path{#1}}}
\providecommand{\bibinfo}[2]{#2}
\ifx\xfnm\relax \def\xfnm[#1]{\unskip,\space#1}\fi
%Type = Article
\bibitem[{Acerbi and Sz{\'e}kely(2014)}]{Acerbi2014}
\bibinfo{author}{Acerbi, C.}, \bibinfo{author}{Sz{\'e}kely, B.},
  \bibinfo{year}{2014}.
\newblock \bibinfo{title}{Back-testing expected shortfall}.
\newblock \bibinfo{journal}{Risk} , \bibinfo{pages}{76--81}.
%Type = Article
\bibitem[{Acerbi and Tasche(2002)}]{Acerbi2002}
\bibinfo{author}{Acerbi, C.}, \bibinfo{author}{Tasche, D.},
  \bibinfo{year}{2002}.
\newblock \bibinfo{title}{On the coherence of expected shortfall}.
\newblock \bibinfo{journal}{Journal of Banking \& Finance}
  \bibinfo{volume}{26}, \bibinfo{pages}{1487--1503}.
%Type = Article
\bibitem[{Bagnato et~al.(2015)Bagnato, Pot{\`\i} and Zoia}]{Bagnato2015a}
\bibinfo{author}{Bagnato, L.}, \bibinfo{author}{Pot{\`\i}, V.},
  \bibinfo{author}{Zoia, M.G.}, \bibinfo{year}{2015}.
\newblock \bibinfo{title}{The role of orthogonal polynomials in adjusting
  hyperpolic secant and logistic distributions to analyse financial asset
  returns}.
\newblock \bibinfo{journal}{Statistical Papers} \bibinfo{volume}{56},
  \bibinfo{pages}{1205--1234}.
%Type = Article
\bibitem[{Balanda and MacGillivray(1988)}]{Balanda1988}
\bibinfo{author}{Balanda, K.P.}, \bibinfo{author}{MacGillivray, H.},
  \bibinfo{year}{1988}.
\newblock \bibinfo{title}{Kurtosis: a critical review}.
\newblock \bibinfo{journal}{The American Statistician} \bibinfo{volume}{42},
  \bibinfo{pages}{111--119}.
%Type = Book
\bibitem[{Bateman(1954)}]{Bateman1954}
\bibinfo{author}{Bateman, H.}, \bibinfo{year}{1954}.
\newblock \bibinfo{title}{Tables of integral transforms [volumes I \& II]}.
  volume~\bibinfo{volume}{1}.
\newblock \bibinfo{publisher}{McGraw-Hill Book Company}.
%Type = Article
\bibitem[{Ding(2014)}]{Ding2014a}
\bibinfo{author}{Ding, P.}, \bibinfo{year}{2014}.
\newblock \bibinfo{title}{{Three occurrences of the Hyperbolic-Secant
  distribution}}.
\newblock \bibinfo{journal}{American Statistician} \bibinfo{volume}{68},
  \bibinfo{pages}{32--35}.
%Type = Article
\bibitem[{Dodd(1925)}]{Dodd1925}
\bibinfo{author}{Dodd, E.L.}, \bibinfo{year}{1925}.
\newblock \bibinfo{title}{{The Frequency Law of a Function of Variables With
  Given Frequency Laws}}.
\newblock \bibinfo{journal}{The Annals of Mathematics} \bibinfo{volume}{27},
  \bibinfo{pages}{12--20}.
%Type = Article
\bibitem[{Finucan(1964)}]{Finucan1964}
\bibinfo{author}{Finucan, H.}, \bibinfo{year}{1964}.
\newblock \bibinfo{title}{A note on kurtosis}.
\newblock \bibinfo{journal}{Journal of the Royal Statistical Society: Series B
  (Methodological)} \bibinfo{volume}{26}, \bibinfo{pages}{111--112}.
%Type = Article
\bibitem[{Fisher(1921)}]{Fisher1921}
\bibinfo{author}{Fisher, R.A.}, \bibinfo{year}{1921}.
\newblock \bibinfo{title}{On the probable error of a coefficient of correlation
  deduced from a small sample}.
\newblock \bibinfo{journal}{Metron} \bibinfo{volume}{1},
  \bibinfo{pages}{3--32}.
%Type = Misc
\bibitem[{Graham et~al.(1994)Graham, Knuth and Patashnik}]{Graham1994}
\bibinfo{author}{Graham, R.L.}, \bibinfo{author}{Knuth, D.E.},
  \bibinfo{author}{Patashnik, O.}, \bibinfo{year}{1994}.
\newblock \bibinfo{title}{Concrete mathematics: A foundation for computer
  science}.
%Type = Book
\bibitem[{Gut(2009)}]{Gut2013}
\bibinfo{author}{Gut, A.}, \bibinfo{year}{2009}.
\newblock \bibinfo{title}{An Intermediate Course in Probability}.
\newblock \bibinfo{publisher}{Springer}.
%Type = Article
\bibitem[{Joe and Xu(1996)}]{Joe1996}
\bibinfo{author}{Joe, H.}, \bibinfo{author}{Xu, J.J.}, \bibinfo{year}{1996}.
\newblock \bibinfo{title}{{The Estimation Method of Inference Functions for
  Margins for Multivariate Models}}.
\newblock \bibinfo{journal}{Technical Report no. 166, Department of Statistics,
  University of British Columbia} .
%Type = Article
\bibitem[{Kunsch(1989)}]{Kunsch1989}
\bibinfo{author}{Kunsch, H.R.}, \bibinfo{year}{1989}.
\newblock \bibinfo{title}{The jackknife and the bootstrap for general
  stationary observations}.
\newblock \bibinfo{journal}{The Annals of Statistics} ,
  \bibinfo{pages}{1217--1241}.
%Type = Article
\bibitem[{Kupiec(1995)}]{Kupiec1995}
\bibinfo{author}{Kupiec, P.}, \bibinfo{year}{1995}.
\newblock \bibinfo{title}{Techniques for verifying the accuracy of risk
  measurement models}.
\newblock \bibinfo{journal}{The Journal of Derivatives} \bibinfo{volume}{3}.
%Type = Article
\bibitem[{McNeil and Frey(2000)}]{McNeil2000}
\bibinfo{author}{McNeil, A.J.}, \bibinfo{author}{Frey, R.},
  \bibinfo{year}{2000}.
\newblock \bibinfo{title}{Estimation of tail-related risk measures for
  heteroscedastic financial time series: an extreme value approach}.
\newblock \bibinfo{journal}{Journal of empirical finance} \bibinfo{volume}{7},
  \bibinfo{pages}{271--300}.
%Type = Article
\bibitem[{Morris(1982)}]{Morris1982}
\bibinfo{author}{Morris, C.N.}, \bibinfo{year}{1982}.
\newblock \bibinfo{title}{Natural exponential families with quadratic variance
  functions}.
\newblock \bibinfo{journal}{The Annals of Statistics} \bibinfo{volume}{10},
  \bibinfo{pages}{65--80}.
%Type = Article
\bibitem[{Morris and Lock(2009)}]{Morris2009}
\bibinfo{author}{Morris, C.N.}, \bibinfo{author}{Lock, K.F.},
  \bibinfo{year}{2009}.
\newblock \bibinfo{title}{{Unifying the named natural exponential families and
  their relatives}}.
\newblock \bibinfo{journal}{American Statistician} \bibinfo{volume}{63},
  \bibinfo{pages}{247--253}.
%Type = Book
\bibitem[{Nelsen(2007)}]{Nelsen2007}
\bibinfo{author}{Nelsen, R.B.}, \bibinfo{year}{2007}.
\newblock \bibinfo{title}{An introduction to copulas}.
\newblock \bibinfo{publisher}{Springer Science \& Business Media}.
%Type = Article
\bibitem[{Nyblom(1999)}]{Nyblom1999}
\bibinfo{author}{Nyblom, M.}, \bibinfo{year}{1999}.
\newblock \bibinfo{title}{On a generalization of the binomial theorem}.
\newblock \bibinfo{journal}{The Fibonacci Quarterly} \bibinfo{volume}{37},
  \bibinfo{pages}{3--13}.
%Type = Article
\bibitem[{Perks(1932)}]{Perks1932}
\bibinfo{author}{Perks, W.}, \bibinfo{year}{1932}.
\newblock \bibinfo{title}{On some experiments in the graduation of mortality
  statistics}.
\newblock \bibinfo{journal}{Journal of the Institute of Actuaries}
  \bibinfo{volume}{63}, \bibinfo{pages}{12--57}.
%Type = Book
\bibitem[{Roa(1924)}]{Roa1924}
\bibinfo{author}{Roa, E.}, \bibinfo{year}{1924}.
\newblock \bibinfo{title}{A number of new generating functions with
  applications to statistics}.
\newblock \bibinfo{publisher}{Lancaster Press}.
%Type = Article
\bibitem[{Varin et~al.(2011)Varin, Reid and Firth}]{Varin2011}
\bibinfo{author}{Varin, C.}, \bibinfo{author}{Reid, N.M.},
  \bibinfo{author}{Firth, D.}, \bibinfo{year}{2011}.
\newblock \bibinfo{title}{{An overview of composite likelihood methods}}.
\newblock \bibinfo{journal}{Statistica Sinica} \bibinfo{volume}{21},
  \bibinfo{pages}{5--42}.
%Type = Article
\bibitem[{Vaughan(2002)}]{Vaughan2002}
\bibinfo{author}{Vaughan, D.C.}, \bibinfo{year}{2002}.
\newblock \bibinfo{title}{The generalized secant hyperbolic distribution and
  its properties}.
\newblock \bibinfo{journal}{Communications in Statistics-Theory and Methods}
  \bibinfo{volume}{31}, \bibinfo{pages}{219--238}.
%Type = Article
\bibitem[{Zoia et~al.(2018)Zoia, Biffi and Nicolussi}]{Zoia2018}
\bibinfo{author}{Zoia, M.G.}, \bibinfo{author}{Biffi, P.},
  \bibinfo{author}{Nicolussi, F.}, \bibinfo{year}{2018}.
\newblock \bibinfo{title}{{Value at risk and expected shortfall based on
  Gram-Charlier-like expansions}}.
\newblock \bibinfo{journal}{Journal of Banking and Finance}
  \bibinfo{volume}{93}, \bibinfo{pages}{92--104}.

\end{thebibliography}
\end{document}